\documentclass[11pt,twoside]{article}
\usepackage{amsmath}
\usepackage{amssymb}
\usepackage{mathrsfs}
\usepackage{graphicx}
\usepackage{epstopdf}
\newtheorem{corollary}{Corollary}[section]
\newtheorem{definition}{Definition}[section]
\newtheorem{lemma}{Lemma}[section]
\newtheorem{proposition}{Proposition}[section]
\def\EndBox#1{
	\hskip0.1em\hfill\null\ \null\nobreak\hfill\kern3pt
		\hbox{$\scriptstyle #1$} \smallbreak}
\def\qed{\EndBox{\square}}
\newcommand{\proof}{{\sc proof:~}}
\newcommand{\remark}{\smallbreak\noindent{\bf Remark.}}
\renewcommand{\a}{\alpha}
\renewcommand{\b}{\beta}
\newcommand{\g}{\gamma}
\renewcommand{\d}{\delta}
\newcommand{\e}{\varepsilon}
\newcommand{\ep}{\epsilon}
\newcommand{\ze}{{\zeta}}
\renewcommand{\th}{\theta}
\newcommand{\vth}{\vartheta}
\newcommand{\io}{\iota}
\renewcommand{\l}{\lambda}
\newcommand{\m}{\mu}
\newcommand{\n}{\nu}
\renewcommand{\r}{\rho}
\newcommand{\s}{\sigma}

\renewcommand{\t}{\tau}
\newcommand{\vph}{\varphi}
\newcommand{\om}{{\omega}}
\newcommand{\h}{\hbar}
\newcommand{\de}{\partial}
\newcommand{\Eo}{{\scriptstyle{\mathrm{E}}}}
\newcommand{\oh}{\tfrac{1}{2}}
\newcommand{\ih}{\tfrac{\iO}{2}}
\newcommand{\INT}[1]{{\!\!\!\rlap{{\lower6pt\hbox{$
			\begin{array}{l}{\displaystyle\int}\\[-2pt]
			{{\scriptstyle#1}}\end{array}$}}}}\hspace{17pt}}
\newcommand{\Dsl}[1]{{\rlap{\raise1pt\hbox{\,/}}#1}}
\newcommand{\dAl}{\,{\begin{picture}(7,7)
  \put(-.75,0){$\square$}\put(0,-.22){\line(1,0){7.5}}
  \put(7.33,-.25){\line(0,1){7.54}}\end{picture}\,}}
\newcommand{\cj}[1]{\overline{#1}}
\newcommand{\td}{\tilde}
\newcommand{\one}{{\scriptscriptstyle{(1)}}}
\newcommand{\two}{{\scriptscriptstyle{(2)}}}
\newcommand{\cev}[1]{\smash{\overset{\smash{{}_{\gets}}}{#1}}}
\newcommand{\free}{{}_{\sst\mathrm{free}}}
\newcommand{\sbot}{{\scriptscriptstyle\bot}}
\newcommand{\sbo}{{\!\sbot}}
\newcommand{\spar}{{\scriptscriptstyle\|}}
\newcommand{\adv}{{\scriptscriptstyle\mathrm{adv}}}
\newcommand{\ret}{{\scriptscriptstyle\mathrm{ret}}}
\newcommand{\Fey}{{\scriptscriptstyle\mathrm{F}}}
\newcommand{\ost}[1]{\overset{{}_{{\,}_*}}{#1}}
\newcommand{\C}{{\boldsymbol{C}}}
\newcommand{\E}{{\boldsymbol{E}}}
\newcommand{\K}{{\boldsymbol{K}}}
\newcommand{\M}{{\boldsymbol{M}}}
\newcommand{\Mm}{{\scriptscriptstyle{\boldsymbol{M}}}}
\newcommand{\N}{{\boldsymbol{N}}}
\renewcommand{\P}{{\boldsymbol{P}}}
\newcommand{\Pm}{\P_{\!\!m}}
\newcommand{\U}{{\boldsymbol{U}}}
\newcommand{\Uu}{{\scriptscriptstyle{\boldsymbol{U}}}}
\newcommand{\V}{{\boldsymbol{V}}}
\newcommand{\W}{{\boldsymbol{W}}}
\newcommand{\Ww}{{\scriptscriptstyle{\boldsymbol{W}}}}
\newcommand{\X}{{\boldsymbol{X}}}
\newcommand{\Xx}{{\scriptscriptstyle{\boldsymbol{X}}}}
\newcommand{\Z}{{\boldsymbol{Z}}}
\newcommand{\Zc}{\cj{\Z}}
\newcommand{\CC}{{\mathbb{C}}}
\newcommand{\KK}{{\mathbb{K}}}
\newcommand{\LL}{{\mathbb{L}}}
\newcommand{\MM}{{\mathbb{M}}}
\newcommand{\NN}{{\mathbb{N}}}
\newcommand{\QQ}{{\mathbb{Q}}}
\newcommand{\RR}{{\mathbb{R}}}
\newcommand{\SSI}{{\mathbb{S}}}
\newcommand{\TT}{{\mathbb{T}}}
\newcommand{\UU}{{\mathbb{U}}}
\newcommand{\VV}{{\mathbb{V}}}
\newcommand{\ZZ}{{\mathbb{Z}}}
\newcommand{\Acal}{{\mathcal{A}}}
\newcommand{\Dcal}{{\mathcal{D}}}
\newcommand{\Ecal}{{\mathcal{E}}}

\newcommand{\Hcal}{{\mathcal{H}}}
\newcommand{\Lcal}{{\mathcal{L}}}
\newcommand{\Ocal}{{\mathcal{O}}}
\newcommand{\Qcal}{{\mathcal{Q}}}
\newcommand{\Scal}{{\mathcal{S}}}
\newcommand{\Ucal}{{\mathcal{U}}}
\newcommand{\Vcal}{{\mathcal{V}}}
\newcommand{\Zcal}{{\mathcal{Z}}}
\newcommand{\AC}{{\boldsymbol{\Acal}}}
\newcommand{\DC}{{\boldsymbol{\Dcal}}}
\newcommand{\EC}{{\boldsymbol{\Ecal}}}
\newcommand{\HC}{{\boldsymbol{\Hcal}}}
\newcommand{\LC}{{\boldsymbol{\Lcal}}}
\newcommand{\OC}{{\boldsymbol{\Ocal}}}
\newcommand{\QC}{{\boldsymbol{\Qcal}}}
\newcommand{\SC}{{\boldsymbol{\Scal}}}
\newcommand{\UC}{{\boldsymbol{\Ucal}}}
\newcommand{\VC}{{\boldsymbol{\Vcal}}}
\newcommand{\ZC}{{\boldsymbol{\Zcal}}}
\newcommand{\DCo}{\DC_{\!\circ}}
\newcommand{\DCv}{\DC_{\!_\VV}}
\newcommand{\DCh}{{\rlap{\;/}\DC}}
\newcommand{\DCho}{{\rlap{\;/}\DCo}}
\newcommand{\QCo}{\QC_\circ}
\newcommand{\SCo}{\SC_{\!\circ}}
\newcommand{\SCv}{\SC_{\!_\VV}}
\newcommand{\VCo}{\VC_{\!\circ}}
\newcommand{\ZCo}{\ZC_{\!\circ}}

\newcommand{\supp}{\operatorname{supp}}
\newcommand{\End}{\operatorname{End}}

\newcommand{\pv}{\operatorname{pv}}
\newcommand{\sgn}{\operatorname{sgn}}
\newcommand{\Id}[1]{{1\!\!1}\!{}_{#1}{}}
\newcommand{\id}{{1\!\!1}}
\newcommand{\dO}{\mathrm{d}}
\newcommand{\DO}{\mathrm{D}}
\newcommand{\FO}{\mathrm{F}}
\newcommand{\LO}{\mathrm{L}}
\newcommand{\he}{{\scriptstyle{\mathrm{H}}}}
\newcommand{\TO}{\mathrm{T}}
\newcommand{\TS}{\TO^{*}\!}
\newcommand{\dth}{\dO\th}
\newcommand{\dph}{\dO\phi}

\newcommand{\drh}{\dO\r}
\newcommand{\dvth}{\dO\vth}
\newcommand{\dvph}{\dO\vph}
\newcommand{\eO}{\mathrm{e}}
\newcommand{\iO}{\mathrm{i}}
\newcommand{\na}{\nabla\!}
\newcommand{\nasl}{{\rlap{\raise1pt\hbox{\,/}}\nabla}}
\newcommand{\ten}[1]{\operatorname*{\otimes}_{\!{\scriptscriptstyle #1}} }
\newcommand{\we}{{\,\wedge\,}}
\newcommand{\weu}[1]{{\wedge^{\!#1}}}
\newcommand{\mdots}{{\cdot}{\cdot}{\cdot}}
\newcommand{\pint}{\mathord{\rfloor}}
\newcommand{\comp}{\mathbin{\raisebox{1pt}{$\scriptstyle\circ$}}}
\newcommand{\tn}{{\,\otimes\,}}
\newcommand{\tnb}{{\,\smash{\stackrel{{}_{{\centerdot}}}{\otimes}}\,}}

\newcommand{\Bang}[1]{{\left\langle#1\right\rangle}}
\newcommand{\bang}[1]{{\langle#1\rangle}}
\newcommand{\bbang}[1]{{\pmb{\langle}#1\pmb{\rangle}}}
\newcommand{\ket}[1]{{|#1\rangle}}
\newcommand{\sA}{{\scriptscriptstyle A}}
\newcommand{\Bsf}{{\mathsf{B}}}
\newcommand{\bb}{{\mathsf{b}}}
\newcommand{\Ssf}{{\mathsf{S}}}
\newcommand{\Xsf}{{\mathsf{X}}}
\newcommand{\ff}{{\mathsf{f}}}
\newcommand{\pp}{{\mathsf{p}}}
\newcommand{\p}{{\scriptstyle\Pi}}
\newcommand{\sref}[1]{\S\ref{#1}}
\newcommand{\ie}{i.e$.$}
\newcommand{\sst}{\scriptscriptstyle}
\newcommand{\gto}{\leadsto}
\newcommand{\into}{\hookrightarrow}
\newcommand{\onto}{\rightarrowtail}
\newcommand{\ul}{\underline}
\newcommand{\opp}[3]{o_{#1}\raise1.3pt\hbox{$\scriptscriptstyle[#2#3]$}}
\newcommand{\vpp}[3]{\check{o}_{#1}\raise1.3pt\hbox{$\scriptscriptstyle[#2#3]$}}
\newcommand{\fpp}[3]{\hat{o}_{#1}\raise1.3pt\hbox{$\scriptscriptstyle[#2#3]$}}
\newcommand{\abs}[1]{{\mathrm{#1}}}
\newcommand{\tra}[1]{\abs{#1}^{\!*}}
\newcommand{\suc}[1]{{\{\![#1]\!\}}}
\newcommand{\Suc}[1]{{\bigl\{\!\!\bigl[#1\bigr]\!\!\bigl\}}}
\newcommand{\swe}{\,{\scriptstyle\lozenge}\,}
\newcommand{\grade}[1]{{\lfloor#1\rceil}}
\title{Special generalized densities and propagators:\\ a geometric account}
\date{{\small v2: September 2, 2015} }
\author{Daniel Canarutto\\[6pt]
{\small\it Dipartimento di Matematica e Informatica ``U.~Dini'', }\\
{\small\it Via S. Marta 3, 50139 Firenze, Italia}\\
{\small email:~daniel.canarutto@unifi.it}\\
{\small http://www.dma.unifi.it/\char126 canarutto}}
\begin{document}
\bibliographystyle{alpha}
\maketitle \thispagestyle{empty}
\begin{abstract}\noindent
Starting from a short review of spaces of generalized sections of vector bundles,
we give a concise systematic description, in precise geometric terms,
of Leray densities, principal value densities,
propagators and elementary solutions of field equations in flat spacetime.
We then sketch a partly original geometric presentation of free quantum fields
and show how propagators arise from their graded commutators
in the boson and fermion cases.
\end{abstract}

\bigbreak

\bigbreak

\vfill\newpage
\tableofcontents
\thispagestyle{empty}
\vfill\newpage
\setcounter{page}{1}

\section*{Introduction}

Propagators and delta-functions are ubiquitous
in the physiscs literature~\cite{IZ,We96},
but are often introduced in an informal way by writing down
the main needed formulae and then reducing other calculations to those,
possibly by \emph{ad hoc} variable changes.
However it's reasonable to say that the matter can be better handled
if we avail of a more systematic treatment,
based on a precise geometric description in terms of distributions.
These are the subject of a beautiful and rigorous mathematical theory,
covered by excellent classical texts.
The use of the fundamental notions of this theory
is quite natural once they are assimilated,
though at first sight the most abstract concepts, and the proofs of the main theorems,
may not appeal to the practical-minded.
We could even argue that generalized densities
are more natural in physics than actual functions.

Section~\ref{s:Distributional spaces} aims at a concise
but hopefully clear enough summary of notions and results
about spaces of generalized sections of vector bundles.
While our notation is not completely standard,
and we also focus on a few concepts
which are not often treated explicitely,
the essentials of most details and proofs can be found, for example,
in the classical text by Schwartz~\cite{Sc},
or in Choquet-Bruhat~and~DeWitt-Morette~\cite{CdW}.
We specially insist on the geometric aspect of the matter.
We consider generalized  densities, semi-densities and currents,
as well as delta-type densities and principal value densities.

In section~\ref{s:Special generalized densities on Minkowski spacetime}
we try to give a systematic though synthetic introduction to Leray densities,
principal value densities, propagators and elementary solutions
of field equations in flat spacetime.
We aim at a thorough exploration
of the relations among several involved objects,
the different ways in which they can be defined and their Fourier transforms,
as well as at their roles in the determination of
elementary solutions of the Klein-Gordon equation and of the Dirac equation
(with the D'Alembert equation and the Weyl equation as special cases).

In section~\ref{s:Graded commutators of quantum fields}
we sketch a partly original geometric presentation of free quantum fields
introduced in previous papers~\cite{C14b,C15a},
and show how propagators arise from their graded commutators
both in the boson and in the fermion case.

\section{Distributional spaces}\label{s:Distributional spaces}
\subsection{Unit spaces}\label{ss:Unit spaces}

Physical scales (or ``dimensions'') are usually dealt with in an informal way,
but a more formal approach helps to clarify the geometric background of a theory.
Now and then various proposals pop up
in the literature~\cite{Carlson79,Tao12,Benioff15},
often apparently unaware of each other,
but we maintain that by far the clearest and algebraically most precise setting
was introduced around 1995 after an idea of M.\:Modugno,
and has been used since then
in papers of various authors.\footnote{
See e.g.\ \cite{JMV10,C12a} and the bibliography therein.} 
In particular, the notion of \emph{1-dimensional positive space}
captures the essential point about unit spaces,
and tensor products between a unit space and a vector space allow us
to handle spaces which are similar but differently scaled.
We also stress that the ensuing formalism is quite handy,
though the demonstrations of certain basic properties
are not trivial~\cite{JMV10}.

We now sketch the basic notions to be used in the sequel.

A \emph{semi-vector space} is defined to be
a set equipped with an internal addition map
and multiplication by positive reals,
fulfilling the usual axioms of vector spaces
except those properties which involve opposites and the zero element.
A semi-vector space $\UU$ is called a \emph{positive space}
if the multiplication \hbox{$\RR^+\,{\times}\,\UU\to\UU$} is a transitive
left action of the group $(\RR^+,{\cdot}\,)$ on $\UU$.
Then a positive space cannot have a zero element.
In particular, any vector space is a semi-vector space,
while the set of linear combinations over $\RR^+$ of $n$ independent vectors
in a vector space is a positive space.

Most algebraic notions related to vector spaces,
like dimensionality, linearity, duality and tensor products,
can be seamlessly extended to positive spaces.
A $1$-dimensional positive space is called a \emph{unit space},
or a \emph{scale space}.
Integer tensor powers \hbox{$\UU^p\equiv\UU\tn\mdots\tn\UU$}
(\hbox{$p\in\NN$} factors)
of a unit space $\UU$ are again unit spaces.
This also holds for negative tensor powers,
defined as \hbox{$\UU^{-p}\equiv(\UU^*)^p$}.
Moreover if $\V$ is a vector space then the tensor product $\UU\tn\V$
is a well-defined vector space which can be regarded
as a ``scaled'' version of $\V$.
If \hbox{$(u,v)\in\UU\times\V$} then we use the ``number-like'' notation
\hbox{$u\,v\equiv u\tn v$}.
Moreover $u$ has a unique inverse or dual \hbox{$u^{-1}\equiv u^*\in\UU^{-1}$},
so that we can \emph{formally} treat elements in $\UU$ as positive numbers.

Furthermore for any \hbox{$p\in\NN$} there is a natural construction~\cite{C12a}
of a unit space $\UU^{1/p}$ fulfilling \hbox{$(\UU^{1/p})^p\cong\UU$},
hence we also obtain rational powers of unit spaces.

In many physical theories it is convenient to assume
the space $\TT$ of \emph{time units},
the space $\LL$ of \emph{length units},
and the space $\MM$ of \emph{mass units},
and construct any other needed scale space as
\hbox{$\SSI=\TT^{d_1}\tn\LL^{d_2}\tn\MM^{d_3}$} with \hbox{$d_i\in\QQ$}\,.
Two sections \hbox{$\s:\M\to\E$} and \hbox{$\s':\M\to\SSI\tn\E$}
of differently scaled vector bundles
can be compared by means of a \emph{coupling constant} \hbox{$s\in\SSI$}.
In an unscaled frame, the components of an $\SSI$-scaled section are
valued into $\SSI\tn\RR$\,.
In particular we have the {speed of light} \hbox{$c\in\TT^{-1}\tn\LL$}
and \emph{Planck's constant} \hbox{$\h\in\TT^{-1}\tn\LL^2\tn\MM$}\,.
Together, these determine isomorphisms
\hbox{$\TT\cong\LL$} and \hbox{$\MM\cong\LL^{-1}$},
so we can actually reduce all scale spaces to powers of $\LL$\,;
this is sometimes called the \emph{natural} system,
corresponding to the setting \hbox{$c=\h=1$}.
Then, in particular, a \emph{mass} is an element \hbox{$m\in\LL^{-1}$} and,
in Einstein spacetime $(\M,g)$,
we identify the bundle \hbox{$\P\cong\MM\tn\LL^2\tn\TT^{-1}\tn\TS\M$}
of $4$-momenta with $\TS\M$.

\subsection{Generalized sections}\label{ss:Generalized sections}

If $\M$ is an $m$-dimensional oriented classical manifold\footnote{
By a \emph{classical manifold} we mean a Hausdorff paracompact
smooth real manifold of finite dimension
(while typical quantum state spaces and bundles
are infinite-dimensional).} 
without boundary, then the fibers of the vector bundle
$$\VV\M\equiv(\weu{m}\TO\M)^{+}\onto\M$$
are unit spaces.
We call \hbox{$\VV^*\M\equiv(\weu{m}\TS\M)^{+}\onto\M$}
the bundle of all \emph{positive densities} (or \emph{volume forms}) on $\M$.
If there is no danger of confusion we use the shorthand \hbox{$\VV\equiv\VV\M$}.

Let \hbox{$\pp:\V\onto\M$} be an either real or complex
finite-dimensional vector bundle,
and denote by $\DCo(\M,\V)$ the vector space of all global smooth sections
\hbox{$\M\to\V$} which have compact support, called \emph{test sections}.
If $\X\subset\M$ is an open subset, then we have a natural vector subspace
$$\DCo(\X,\V_{\!\!\!\Xx})\subset\DCo(\M,\V)~,
$$
where \hbox{$\V_{\!\!\!\Xx}\equiv\cev\pp(\X)$} is the preimage of $\X$
through $\pp$\,.
In fact any compact subset in $\X$ is also a compact subset in $\M$,
and the natural smooth extension of an element in $\DCo(\X,\V_{\!\!\!\Xx})$
is obtained by letting it vanish in \hbox{$\M\setminus\X$}.
However note that if \hbox{$\K\subset\M$} is compact
then \hbox{$\K\cap\X$} is not compact in general,
so there is no natural projection
\hbox{$\DCo(\M,\V)\onto\DCo(\X,\V_{\!\!\!\Xx})$}\,.

\begin{definition}\label{d:testmaptopology1}
A sequence \hbox{$(u_j):\NN\to\DCo(\M,\V)$} is said to be
\emph{converging to zero}, or a \emph{null sequence}, if
for any fibered coordinate chart\,\footnote{
\,$\KK$ is either $\RR$ or $\CC$
} 
\hbox{$\V_{\!\!\!\Xx}\to\RR^m\times\KK^n$}
its coordinate expression
and the coordinate expressions of its partial derivatives, at any order,
all converge uniformly to $0$ on every compact subset \hbox{$\C\subset\X$}.
\end{definition}
By considering a bundle atlas on \hbox{$\V\onto\M$}
one obtains\footnote{
See e.g.\ Schwartz~\cite{Sc}, in particular \S\,III.1 and \S\,IX.2.} 
a topology on $\DCo(\M,\V)$.
A linear map \hbox{$\phi:\DCo(\M,\V)\to\KK$} is continuous if
$$\lim_{j\to\infty}\phi(u_j)=0$$
for every null sequence \hbox{$(u_j):\NN\to\DCo(\M,\V)$}\,.
The vector space of all such continuous functionals,
also called \emph{distributions},
is the topological dual of $\DCo(\M,\V)$,
which we denote as\footnote{In the standard literature,
the most usual notation is $\DC$ for the space of test maps
and $\DC'$ for the corresponding distributional space.}
$$\DCv(\M,\V^*)\equiv\DC(\M,\VV^*\ten{\M}\V^*)~.$$
This can be seen as a space of \emph{generalized sections},
in the following sense.
We define an ordinary $\V^*$-valued density\footnote{
Here \hbox{$\V^{*}\onto\M$} may stand
for either the real or the complex dual bundle.} 
\hbox{$\th:\M\to\VV^*\ten{\M}\V^*$}
to be \emph{locally integrable} if the ordinary density
$$\bang{\th,u}:\M\to\KK\tn\VV^{*}:x\mapsto\bang{\th(x),u(x)}$$
is integrable for all \hbox{$u\in\DCo(\M,\V)$}\,.
Then $\th$ can be identified with a distribution by the rule
$$\th(u)\equiv\bbang{\th,u}:=\int_\Mm \bang{\th,u}~.$$
Hence any \hbox{$\th\in\DCv(\M,\V^*)$} is called
a \emph{generalized $\V^*$-valued density}.
Accordingly, we'll also write \hbox{$\th:\M\gto\VV^*\ten{\M}\V^*$}.

Moreover $\DCv(\M,\V^*)$ has a natural topology~\cite{Sc},
such that every element $\th$ of its can be written as the limit, in a strong sense,
of sequences $(\th_j)$ of ordinary locally integrable densities.
This implies also weak convergence,
namely for any \hbox{$u\in\DCo(\M,\V)$} we have
$$\bbang{\th,u}=\lim_{j\to\infty}\bbang{\th_j\,,u}
=\lim_{j\to\infty}\int_\Mm \bang{\th_j,u}~.$$
This justifies the use of writing $\bbang{\th,u}$ in the form of an ordinary integral
(to be intended in a \emph{generalized sense}),
even if the value $\th(x)$ at any $x\in\M$ may have no meaning at all.
Also, the formal properties of the integral still hold.

The vector space $\DCv(\M,\V^*)$
has the vector subspace $\DCo(\M,\VV^*\!\M\ten{\M}\V^*)$
of all ordinary smooth $\V^*$-valued densities with compact support,
whose topological dual is the space $\DC(\M,\V)$
of all \emph{generalized sections} \hbox{$\M\gto\V$}.
Moreover \hbox{$\DCo(\M,\V)\subset\DC(\M,\V)$} is a vector subspace.
So the symbols $\DCo$ and $\DC$ can be used for test sections
and generalised sections of any vector bundle over $\M$;
the shorthand $\DCv$ is useful for denoting generalized densities.

It turns out that $\DCo(\M,\V)$ is dense in $\DC(\M,\V)$, namely
for every \hbox{$\s\in\DC(\M,\V)$} there exist sequences
\hbox{$(\s_j):\NN\to\DCo(\M,\V)$} converging to $\s$\,.

The natural inclusion
\hbox{$\DCo(\X,\VV^*\!\X\ten{\X}\V^*_{\!\!\!\Xx})
\subset\DCo(\M,\VV^*\!\M\ten{\M}\V^*)$},
where \hbox{$\X\subset\M$} is any open subset,
determines the natural projection
\hbox{$\DC(\M,\V)\onto\DC(\X,\V_{\!\!\!\Xx})$}
given by restriction.
On the other hand,
there is no natural inclusion \hbox{$\DC(\X,\V_{\!\!\!\Xx})\subset\DC(\M,\V)$},
as there is no natural way of extending a distribution on $\X$
(even more, such extension may not exist at all).
However, a gluing property holds:
if $\{\X\!_\a\}$ is an open covering of $\M$,
\hbox{$\V\!_\a:=\cev\pp(\X_\a)$}
and \hbox{$\{\th_i\in\DC(\X\!_\a\,,\V\!_\b)\}$} is a family of generalized sections
such that $\th_\a$ and $\th_\b$ coincide on \hbox{$\X\!_\a\cap\X\!_\b$}
whenever this is non-empty,
then there is a unique \hbox{$\th\in\DC(\M,\V)$} whose restriction to $\X\!_\a$
coincides with $\th_\a$ for all $\a$\,.

A generalized section has a coordinate expression
just like an ordinary section.
If $\bigl(\bb_i\bigr)$ is a local frame of $\V$
and $\bigl(\bb^i\bigr)$ is the dual frame of $\V^*$,
then \hbox{$\s\in\DC(\M,\V)$} can be locally written
as \hbox{$\s=\s^i\,\bb_i$} where \hbox{$\s^i\in\DC(\M,\KK)$}
is given by
$$\bbang{\s^i,f}\equiv\bbang{\s,f\,\bb^i}~,\quad
\forall\;f\in\DCo(\M,\CC\tn\VV^*)~.$$

Pull-backs and push-forwards of ordinary sections,
relatively to bundle morphisms with appropriate properties,
can be naturally extended to generalized sections.
In particular, a fibred linear automorphism \hbox{$\Phi:\V\to\V$}
over a diffeomorphism of the base manifold
yields a linear automorphism $\Phi_{*}$ of $\DCv(\M,\V)$\,.
Note that in texts in analysis $\Phi_{*}\th$ is sometimes indicated
by $\th\comp\Phi$\,, which is somewhat misleading.
Actually if \hbox{$\th\in\DCv(\RR^m,\CC)$} and \hbox{$F:\RR^m\to\RR^m$}
is continuous and invertible then we have
$$\bbang{F\!_*\th,u}=\bbang{\th,\det\cev{F}\,(u\comp\cev{F})}~,$$

The main operators of standard differential geometry,
like exterior differential, Lie derivative along a vector field,
and the covariant derivative with respect to a given connection,
can be straightforwardly extended to act on the appropriate generalized sections.
The coordinate expressions of such extended operators are formally the same
as those of their standard counterpart.
The constructions are based on integration by parts
and the fact that test sections have compact support.
In particular, if \hbox{$v:\M\to\TO\M$} is a smooth vector field and
\hbox{$f:\M\gto\CC$} is a \emph{generalized function},
then the generalized function $v.f$ is defined by the
requirement that the identity
\hbox{$\bbang{v.f\,,\,\b}=-\bbang{f\,,\,\LO_v\b}$}
holds for any test density $\b$
(this also includes partial derivatives).

\subsection{Generalized currents and semi-densities}
\label{ss:Generalized currents and semi-densities}

The ``target'' $\V$ in $\DC(\M,\V)$ can be replaced by any vector bundle over $\M$.
In particular we obtain the space \hbox{$\DC(\M,\weu{p}\TS\M\ten{\M}\V)$}
of \emph{generalized $\V$-valued $p$-currents} ($p$ is a positive integer),
which by obvious isomorphisms can be seen as the topological dual of the space
\hbox{$\DCo(\M,\weu{m{-}p}\TS\M\ten{\M}\V^*)$}
of test $\V^*$-valued $m{-}p$-currents.
Also note that the assignment of a volume form \hbox{$\eta:\M\to\VV^*\M$}
determines an isomorphism
\hbox{$\DCv(\M,\V)\leftrightarrow\DC(\M,\V)$}\,,
as a generalized density $\om$ can be identified
with the generalized section $\breve\om$ through the relation
\hbox{$\om=\eta\tn\breve\om$}\,.

Since the fibers of \hbox{$\VV\onto\M$} are unit spaces
we can also consider spaces $\DC(\M,\VV^q\ten{\M}\V)$ with \hbox{$q\in\QQ$}\,.
For \hbox{$q=-1/2$} we obtain the space
of generalized $\V$-valued \emph{semi-densities}.
In this special case we use the notation
$$\DCh(\M,\V)\equiv\DC(\M,\VV^{-1/2}\ten{\M}\V)~,
\quad \DCho(\M,\V)\equiv\DCo(\M,\VV^{-1/2}\ten{\M}\V)~.$$
Note how $\DCho(\M,\V)$ is a vector subspace of  $\DCh(\M,\V)$,
while in general the inclusion of an arbitrary test space
into its topological dual is subjected to the choice of a volume form.

Let now assume that \hbox{$\V\onto\M$} is a complex bundle
whose fibers are endowed with a Hermitian structure.
Then we have an anti-isomorphism
\hbox{$\DCho(\M,\V)\to\DCho(\M,\V^*):\l\mapsto\l^\dag$},
which is naturally extended to generalized sections.
\emph{Square-integrable} semi-densities consitute
the space \hbox{$\LC^2\equiv\LC^2(\M,\VV^{-1/2}\ten{\M}\V)$} of all sections
\hbox{$\l:\M\to\VV^{-1/2}\ten{\M}\V$} such that
$$\|\l\|^2:=\int_\Mm |\l|^2\equiv \int_\Mm \bang{\l^\dagger,\l}<\infty~.$$
Let moreover
\hbox{$\tilde{\mathbf0}\equiv\tilde{\mathbf0}(\M,\VV^{-1/2}\ten{\M}\V)$}
be the subspace of all almost-everywhere vanishing semi-densities.
Then
$$\HC\equiv\HC(\M,\VV^{-1/2}\ten{\M}\V):=\LC^2/\tilde{\mathbf0}$$
turns out to be a Hilbert space with the Hermitian product
\hbox{$(\l,\m)\mapsto\bang{\l^\dagger,\m}$}.
We have
$$\DCho\subset\HC\subset\DCh~,$$
namely the triple $(\DCho,\HC,\DCh)$ constitutes a so-called
\emph{rigged Hilbert space}~\cite{BLT}.
Note that $\DCho$ is incomplete in the $\LC^2$ topology and dense in $\HC$
(and also dense in $\DCh$ in the distributional space topology).

\subsection{Tensor products}\label{ss:Tensor products}

The tensor product of any two (possibly infinite-dimensional) vector spaces
$\UC$ and $\VC$ is constituted by all \emph{finite} sums of the type
\hbox{$\sum_i u_i\tn v_i$}\,, with \hbox{$u_i\in\UC$}\,, \hbox{$v_i\in\VC$}\,.
If \hbox{$\V\onto\M$} and \hbox{$\W\onto\N$}
are vector bundles over oriented classical base manifolds then we have
a natural linear inclusion
$$\DCo(\M,\V)\tn\DCo(\N,\W)\into
\DCo\bigl(\M\times\N,\V\ten{\M\times\N}\W\bigr)~,$$
characterized by the identification
\hbox{$(s\tn t)(x,y)\equiv s(x)\tn t(y)$}\,.
It can be proved~\cite{Sc} that the above inclusion is dense,
namely any element in the latter space is the sum of infinite series
of the type \hbox{$\sum_{i=1}^\infty s_i\tn t_i$}\,,
converging in the $\DCo$-topology.
It turns out that,
for any \hbox{$\th\in\DCv(\M,\V)$} and \hbox{$\phi\in\DCv(\N,\W)$},
the ordinary series
\hbox{$\sum_{i=1}^\infty\bbang{\th,s_i} \bbang{\phi,t_i}$}
converges, so that we also have a natural inclusion
$$\DCv(\M,\V)\tn\DCv(\N,\W)\into\DCv\bigl(\M\times\N,\V\ten{\M\times\N}\W\bigr)~.$$
Furthermore this turns out to be a dense inclusion too, namely
any element in the latter space is the sum of infinite series
of the type \hbox{$\sum_{i=1}^\infty \th_i\tn\phi_i$}
converging in the $\DC$-topology.
Then the latter space is the closure of the former,
and we possibly denote it as 
\hbox{$\DCv(\M,\V)\tnb\DCv(\N,\W)$}.

Most standard operations of finite-dimensional tensor algebra,
including contractions, can be naturally extended to the present setting,
though in some cases they may well-defined only in a generalized sense.
We shall deal with such issues, in particular,
in the context of multi-particle state spaces.

Symmetrized and antisymmetrized tensor products
of any given distributional space are straightforwardly introduced
according to the general definition, and naturally extended their closures.
Symmetric and anti-symmetric tensors,
namely elements in \hbox{$\vee^p\DCv(\M,\V)$} or \hbox{$\weu{p}\DCv(\M,\V)$},
\hbox{$p\in\NN$}\,,
can be regarded as either symmetric or anti-symmetric generalized maps
on \hbox{$\M^p\equiv\M\times\mdots\times\M$}.
It may be worthwhile to note that they are \emph{not} valued
in the fiber-symmetrized or antisymmetrized
bundles \hbox{$\vee^p\V\onto\M$} or \hbox{$\weu{p}\V\onto\M$}.

\smallbreak
For notational simplicity we discuss kernels in the case
when there is a distinguished volume form on the base manifold $\M$,
so that one makes the identification \hbox{$\DCv(\M,\V)\cong\DC(\M,\V)$}.
This is true when $\M$ is the spacetime manifold,
which is the main situation of interest in the present paper.
An element \hbox{$K\in\DC(\M,\V^*)\tnb\DC(\M,\V)$}
is called a \emph{kernel} in \hbox{$\V\onto\M$}.
We denote transposition by an asterisk,
\ie\ the transpose kernel is \hbox{$K^*\in\DC(\M,\V)\tnb\DC(\M,\V^*)$}.
One may consider various regularity properties regarding kernels~\cite{Sc}.
In particular, the kernel $K$ is said to be
\emph{left} (resp.\ \emph{right}) \emph{semi-regular} if
for any \hbox{$\l\in\DCo(\M,\V^*)$} (resp.\ \hbox{$u\in\DCo(\M,\V)$})
the generalized section \hbox{$K\pint\l\in\DC(\M,\V^*)$}
(resp.\ \hbox{$u\pint K\in\DC(\M,\V)$}) is actually a smooth section.
A kernel which is both left and right semi-segular is said to be \emph{regular}.

Kernels have an important role in the study of solutions of field equations
(\sref{ss:Elementary solutions of field equations}).

\subsection{Delta-type densities}\label{ss:Delta-type densities}

The spaces $\DC(\M,\KK)$ and $\DCv(\M,\KK)$ of scalar generalized functions
and scalar densities can be seen as special cases of generalized sections
by setting \hbox{$\V\equiv\M\times\KK$}\,.

If \hbox{$\N\subset\M$} is a closed submanifold
then \hbox{$\DCv(\N,\KK)\subset\DCv(\M,\KK)$} is a natural inclusion,
transposed of the linear surjection
\hbox{$\DCo(\M,\KK)\onto\DCo(\N,\KK)$} given by restriction.
Namely, any \hbox{$\n\in\DCv(\N,\KK)$} can be seen
as a generalized density on $\M$ whose support is contained in $\N$.
We may call this a \emph{$\d$-type density}, since the usual Dirac delta
$$\d_{x_0}\equiv\d[x_0]:\DCo(\M,\KK)\to\KK:u\mapsto u(x_0)~,\quad x_0\in\M~,$$
can be seen as a particular case where \hbox{$\N=\{x_0\}$}\,.
Note that not all generalized densities with support contained in $\N$
are in $\DCv(\N,\KK)$.
For example if $v$ is a vector field transversal to $\N$,
and \hbox{$\phi\in\DCv(\N,\KK)$}, then \hbox{$\LO_v\phi\not\in\DCv(\N,\KK)$}\,.

If \hbox{$\n\in\DCv(\N,\KK)\subset\DCv(\M,\KK)$} and \hbox{$s:\M\to\V$}
is an ordinary section,
then we also call \hbox{$\n\tn s\in\DCv(\M,\V)$} a $\d$-type density;
it depends from $s$ only via the values $s$ takes on $\N$.

\begin{proposition}\label{proposition:Lerayexistence}
Let $\M$ be a classical $m$-dimensional manifold,
endowed with a given volume form \hbox{$\eta:\M\to\VV^{*}\!\M$}.
Let \hbox{$f:\M\to\RR$} be a differentiable function such that
the equation \hbox{$f=0$} characterizes a regular submanifold
\hbox{$\io:\N\into\M$} and $\dO f$ never vanishes on $\N$.
Then there exists a unique $m{-}1$-form
$$\om[f]:\N\to\weu{m{-}1}\TS\N$$
characterized by the following property:
for each $m{-}1$-form $\eta':\N\to\weu{m{-}1}\TS\M$,
such that \hbox{$\eta=\dO f\we\eta'$} on $\N$, one has\,\footnote{
See e.g.\ \cite{CdW}, \S\,VI.B.} 
$$\om[f]=\io^*\eta'~.$$
\end{proposition}

\remark~If \hbox{$c\in\RR\setminus\{0\}$} is a constant, then obviously
\hbox{$\om[c\,f]=\frac1c\,\om[f]$}\,.
More generally we can consider scaled functions and densities.
If \hbox{$\eta:\M\to\UU\tn\VV^*\M$} is a scaled volume form
and \hbox{$f:\M\to\RR\tn\SSI$} is a scaled function, then
$$\om[f]:\N\to\UU\tn\SSI^{-1}\tn\weu{m-1}\TS\N~.$$
\smallbreak

The $m{-}1$-form $\om[f]$ of proposition~\ref{proposition:Lerayexistence}
is called a \emph{Leray form}.
The induced generalized density on $\M$ is
denoted by the same symbol.
In the literature it is often denoted by $\d\!_f$ or $\d(f)$\,.
This is explained by the following result,
which is easily proved by means of adapted coordinate charts.
\begin{proposition}\label{p:oneparfamom}
If $\phi_\e$ is a $1$-parameter family of smooth maps \hbox{$\RR\to\RR$},
such that \hbox{$\phi_\e\to\d_0$} for \hbox{$\e\to0$}\,, then
$$\bbang{\om[f],u}=\lim_{\e\to0}\int_{{}_\M} (\phi_\e\comp f)\,u\,\eta~,
\quad u\in\DCo(\M,\KK)~.$$
\end{proposition}

The \emph{Heaviside distribution} $\he$ on $\RR$
is defined by \hbox{$\Bang{\he,u}=\int_0^{+\infty}\!\!u(x)\,\dO x$}
for any test function \hbox{$u\in\DCo(\RR,\RR)$}\,,
and is related to Leray densities by the following result.
\begin{proposition}
If \hbox{$x:\M\to\RR^m$} is a coordinate chart such that \hbox{$f\equiv x^1$} then
$$\om[f]=\tfrac{\de}{\de x^1}(\he\comp f)~.$$
\end{proposition}
Accordingly, if no confusion arises,
we may also write \hbox{$\om[f]=\frac{\de}{\de f}(\he\comp\ff)$}\,.

\smallbreak

If $\U$ is a real vector space and $\eta$ is a volume form on it
then for \hbox{$u_0\in\U$} we write \hbox{$\d_{u_0}=\breve\d_{u_0}\,\eta$},
where $\breve\d_{u_0}$ is the generalized function which is usually denoted
as \hbox{$\d(u\,{-}\,u_0)$}.
We can also introduce the distribution which in particle physics
is denoted by \hbox{$\d(u_1\,{+}\,u_2\,{+}\,\mdots\,{+}\,u_r)$}
as the generalized function on $\U^r$ associated with the Leray density $\om[f]$\,,
with \hbox{$f(u_1\,,\mdots\,,u_r)=u_1\,{+}\,\mdots\,{+}\,u_r$}\,.

\subsection{Principal value densities}\label{ss:Principal value densities}

Let \hbox{$f:\M\to\RR$} and \hbox{$\N\subset\M$} be as in the statement
of proposition~\ref{proposition:Lerayexistence}.
If \hbox{$\th\in\DCv(\M,\V)$} then
\hbox{$\th/f\in\DCv(\M{\setminus}\N,\V)$} is well-defined.
In fact for \hbox{$w\in\DCo(\M{\setminus}\N,\V^*)$}
the support of $w/f$ is bounded away from $\N$,
so that
$$\bbang{\tfrac1f\,\th,w}=\bbang{\th,\tfrac1f\,w}~.$$
It is clear that in general $\th/f$ cannot be extended
to a functional acting on all test sections in $\DCo(\M,\V^*)$,
but in certain cases it is actually possible to find a natural extension.

For \hbox{$\e>0$} let \hbox{$\chi_\e:\M\to\RR$}
be the characteristic function of \hbox{$\M\!_\e:=\cev f(\RR{\setminus}[-\e,\e])$}.
If the limit
$$\pv(\th/f):=\lim_{\e\to0}\bigl(\chi_\e\,\th/f\bigr)$$
exists in $\DCv(\M,\V)$ then it is called
the \emph{(Cauchy) principal value} of $\th/f$\,.

In particular $\pv(\th/f)$ certainly exists if $\th$ is a smooth ordinary density.
This fact can be checked, just for simplicity,
when $\N$ is contained in the domain $\X$
of a coordinate chart $(x^i)$ and \hbox{$f=x^1$}.
If $u\in\DCo(\X,\V^*)$ then
$$\bbang{\pv(\th/f),u}=\lim_{\e\to0}\left(
\int_{-\infty}^{-\e}+\int_\e^{+\infty}\right)\,\dO x^1\,
\INT{\RR^{m-1}}\dO x^2...\,\dO x^m\,\frac{\th^i\,u_i}{x^1}~,$$
which is easily seen to be finite.\footnote{
In the 1-dimensional scalar case, for example, set
\hbox{$\th(x)=\th_0+x\,\th_1(x)$}, \hbox{$u(x)=u_0+x\,u_1(x)$}\,.} 

Note that $\pv(\th/f)$, when it exists,
is \emph{not} the unique extension of $\th/f$\,,
as we can always add a density whose support is in $\N$.
Also, if $\phi$ is an arbitrary density whose support is in $\N$,
then \hbox{$\pv(\phi/f)=0$}\,.
On the other hand $\pv\bigl(\pv(\th/f)/f\bigr)$
is not defined for any smooth $\th$\,,
as this would be the limit for \hbox{$\e\to0$}
of $\th/f^2$ restricted to $\M\!_\e$\,,
which clearly does not exist.

If $f$ and $g$ define regular submanifolds not intersecting each other,
then we have
$$\pv\bigl(\tfrac{f+g}{f\,g}\,\th\bigr)\equiv
\pv\bigl(\tfrac\th f+\tfrac\th g\bigr)=
\pv\bigl(\tfrac\th f\bigr)+\pv\bigl(\tfrac\th g\bigr)~.$$

\subsection{Division}\label{ss:Division}
If \hbox{$\th\in\DCv(\M,\V)$} then the product $f\th$ is well-defined,
for any smooth function \hbox{$f:\M\to\RR$}\,, by
\hbox{$\bbang{f\th\,,\,u}=\bbang{\th\,,\,fu}$}
for any \hbox{$u\in\DCo(\M,\V^*)$}\,.
If $f$ has no zeroes then $\th/f$ is uniquely well-defined too.
In the more general situation when the divisor vanishes somewhere,
we consider the \emph{multiplication equation} \hbox{$f\xi=\th$}
in the unknown $\xi$\,.
\begin{proposition}\label{p:meq0}
If \hbox{$\xi_1\,,\xi_2\in\DCv(\M,\V)$} are solutions of \hbox{$f\xi=\th$}\,,
then their difference is a solution
of the associated homogeneous equation $f\xi=0$\,, and we have
$$\supp(\xi_1-\xi_2)\subset \cev f(0)~.$$
\end{proposition}

\medbreak\noindent
When the divisor vanishes on a regular submanifold we have:
\begin{proposition}\label{p:meq}
Let \hbox{$f:\M\to\RR$} and \hbox{$\N\subset\M$} be as in the statement
of proposition~\ref{proposition:Lerayexistence}.
Then for any \hbox{$\th\in\DCv(\M,\V)$} the equation \hbox{$f\xi=\th$}
has infinitely many solutions \hbox{$\xi\in\DCv(\M,\V)$}\,,
any two solutions differing by an arbitrary element in
\hbox{$\DCv(\N,\V)\subset\DCv(\M,\V)$}\,.
In other terms, the space $\AC$ of all solutions of the above equation
is an affine space, with `derived' vector space \hbox{$\DO\AC=\DCv(\N,\V)$}\,.
\end{proposition}

In particular if $\th$ is a smooth ordinary section then $\pv(\th/f)$ is
a distinguished element in the space $\AC$ of the statement
of the above proposition.
If \hbox{$\th=0$} then \hbox{$\pv(\th/f)=0$} and
we can solve a problem of \emph{vanishing of product}:
the generalized densities $\xi$ fulfilling \hbox{$f\xi=0$}
constitute exactly the space \hbox{$\DCv(\N,\V)\subset\DCv(\M,\V)$}\,.

More complicated situations, in the general division problem,
arise when the considered smooth function has zeroes of order greater than $1$,
or some of its transversal derivatives vanish on \hbox{$\N\equiv\cev f(0)$}\,.

\subsection{Fourier transforms} \label{ss:Fourier transforms}

Throughout this section $\X$ will stand for a real vector space
of finite dimension $m$\,, endowed with a volume form $\eta$\,.
On its dual space $\X^*$ we take the volume form $\ost\eta$
characterized by $(\ost\eta|\eta)=1$\,.
We use linear coordinates $\bigl(x^i\bigr):\X\to\RR^m$ such that
\hbox{$\eta=\dO^mx\equiv\dO x^1\we\mdots\we\dO x^m$}, where \hbox{$\dO x^i=x^i$}.
Similarly we write \hbox{$\ost\eta=\dO^my\equiv\dO y_1\we\mdots\we\dO y_m$}
where $\bigl(y_i\bigr)$ are the dual linear coordinates on $\X^*$.

Let $\SCo(\X,\CC)$ be the topological vector space
of all rapidly decreasing functions \hbox{$\X\to\CC$}~\cite{Sc,CdW}.
Since $\SCo(\X,\CC)$ contains $\DCo(\X,\CC)$ as a dense subspace,
its topological dual $\SCv(\X,\CC)$ is a vector subspace of $\DCv(\X,\CC)$,
called the space of complex \emph{tempered} generalized densities on $\X$.
In particular, any polynomial in the linear coordinates
defines a tempered distribution,
and any distribution with compact support is tempered
(typically, non-tempered distributions grow exponentially at infinity).
It can be proved that $\SCv(\X,\CC)$ is closed under derivation.
By considering the given volume form $\eta$\,,
we can identify it with the space $\SC(\X,\CC)$
of tempered generalized functions.

For each \hbox{$y\in\X^*$} the ``plane wave'' functions
\hbox{$\Ssf^\pm_y:x\mapsto (2\pi)^{-m/2}\,\eO^{\mp\iO\bang{y,x}}$}
are in $\SC(\X,\CC)$,
and it can be seen that if $(u_k)$ is a sequence in $\DCo(\X,\CC)$
converging to \hbox{$u\in\SCo(\X)$}
then the numerical sequences $\bbang{\Ssf^\pm_y,u_k}$ converge.
Then we set
$$\bbang{\Ssf^\pm_y\eta,u}:=\lim_{k\to\infty}\bbang{\Ssf^\pm_y\eta,u_k}~,$$
and obtain maps
$$\FO^\pm u:\X^*\to\CC:y\mapsto\FO^\pm u(y)\equiv\bbang{\Ssf^\pm_y\eta,u}$$
which turn out to be rapidly decreasing.
We obtain continuous linear maps
$$\FO^\pm:\SCo(\X,\CC)\to\SCo(\X^*,\CC):u\mapsto\FO^\pm u$$
which are naturally extended to continuous linear maps
\hbox{$\SC(\X,\CC)\to\SC(\X^*,\CC)$} 
called the \emph{Fourier transform} and \emph{anti-transform}.
Taking the fixed volume form $\eta$ into account,
they can also be seen as linear maps \hbox{$\SCv(\X,\CC)\to\SCv(\X^*,\CC)$}\,.
Actually $\FO^\pm$ turn out to be isomorphisms:
the above constructions and results also hold if we exchange the roles
of $\X$ and $\X^*$,
and then one proves that $\FO^\mp$ is the inverse of $\FO^\pm$.

If $\th$ is a tempered distribution and $u$ is a test map then
$$\bbang{\FO^\pm\th\,,\,u}=\bbang{\th\,,\,\FO^\pm u}~.$$
We'll also use the shorthands
$$\hat\th\equiv\FO^+\th~,\quad \check\th\equiv\FO^-\th~.$$
If $\hat\th$ is an ordinary function then of course
\hbox{$\hat\th(-y)=\check\th(y)$} and the like.

If \hbox{$\th=\breve\th\,\dO^mx$} is an integrable ordinary density,
then for \hbox{$v\in\SCo(\X^*,\CC)$} we have
$$\bbang{\FO^\pm\th,v}=(2\pi)^{-m/2}\,\int_\Xx\int_{\sst{\X^*}}
\breve\th(x)\,v(y)\,\eO^{\mp\iO\bang{y,x}}\,\dO^mx\,\dO^my~.$$

We list a few basic properties for the Fourier transform and anti-trasform
of any tempered distribution $\th$\,. We have
$$\FO^\pm\bigl(\tfrac{\de}{\de x^i}\th\bigr)=\pm\iO\,y_i\,\FO^\mp\th~,\qquad
\FO^\pm\bar\th=\overline{\FO^\mp\th}~,$$
where an overbar denotes complex conjugation.

Let now \hbox{$b\in\X$} and \hbox{$A\in\End\X$}
be a fixed point and a fixed endomorphism of $\X$,
and consider the affine transformation
$$\Phi:\X\to\X:x\mapsto\Phi(x)\equiv b+Ax~.$$
Then we have
$$\FO^\pm(\Phi_*\th)(y)=
\eO^{\mp\iO\bang{y,b}}\,\sgn(\det A)\,(\FO^\pm\th)(A^*y)~,$$
and, in particular,
$$\FO^\pm\bigl(\th(x-b)\bigr)(y)=
\eO^{\mp\iO\bang{y,b}}\,\FO^\pm\th(y)~,
\qquad
\FO^\pm\bigl(\th(ax)\bigr)(y)=|a|^{-m}\,\FO^\pm\th(y/a)\,,~
a\in\RR\setminus\{0\}\,,$$
where we used $x$ as a ``dummy variable''.

We also note that the Fourier transform and anti-trasform
preserve parity symmetry and antisymmetry,
and radial symmetry.
Another important result is the following:
if $\th$ has compact support,
then $\hat\th$ is a smooth slowly increasing function.
Moreover (Paley-Wiener theorem) $\FO^\pm\th$ can be extended to an analytic
map on $\CC\tn\X^*$.
Conversely, the Fourier transform or anti-trasform
of a continuous map fulfilling a suitable bounding condition~\cite{CdW}
is a distribution with compact support.

Finally we note that if $\V$ is a complex vector space
then the Fourier transform and antitransform can be immediately generalized to
inverse isomorphisms
$$\SCv(\X,\V)\stackrel{\phantom{{}^{{}^\pm}}\FO^\pm}{\longleftrightarrow}
\SCv(\X^*,\V)~.$$
If $\th\in\SCv(\X,\V)$ and a basis $(b_i)$ of $\V$ is chosen,
then $\hat\th=\hat\th^i\,b_i$\,, $\check\th=\check\th^i\,b_i$\,.

\subsubsection*{Partial Fourier transforms}
If $\U$ is another vector space then it is natural to introduce the
\emph{partial Fourier transforms and anti-trasforms} by setting
\begin{align*}
&  \FO^\pm_{\!\Uu}:\SCo(\U{\times}\X,\CC)\to\SCo(\U^*{\times}\X,\CC)~,
&& \FO^\pm_{\!\Xx}:\SCo(\U{\times}\X,\CC)\to\SCo(\U{\times}\X^*,\CC)~,
\\[6pt]
& (\FO^\pm_{\!\Uu}w)(v,x):=(\FO^\pm w_x)(v)~,
&&(\FO^\pm_{\!\Xx}w)(u,y):=(\FO w_u)(y)~,
\end{align*}
and extending them to maps acting on temperate distributions via the rules
\begin{align*}
&\bbang{\FO^\pm_{\!\Uu}\th,w}=\bbang{\th,\FO^\pm_{\!\Uu}}~,
&& w\in\SCo(\U^*{\times}\X,\CC)~,\\[6pt]
&\bbang{\FO^\pm_{\!\Xx}\th,w'}:=\bbang{\th,\FO^\pm_{\!\Xx}w'}~,
&& w'\in\SCo(\X{\times}\Z^*,\CC)~.
\end{align*}
In particular we have \hbox{$\FO^+_{\!\Uu}(\l\tn\m)=\hat\l\tn\m$} and the like.
Moreover we have
$$\FO^\pm_{\!\sst{\U\times\X}}=
\FO^\pm_{\!\Uu}\comp\FO^\pm_{\!\Xx}=\FO^\pm_{\!\Uu}\comp\FO^\pm_{\!\Xx}~.$$

\subsubsection*{Fourier transforms: basic examples}

We list a few basic examples.
When needed we treat $x$ as a dummy variable.
Similar formulas can be written by exchanging the roles of $\X$ and $\X^*$.

We have
\begin{align*}
&\FO^\pm(\d[b])(y)=(2\pi)^{-m/2}\,\eO^{\mp\iO\,\bang{y,b}}\,\ost\eta~,\quad b\in\X~,
\\[6pt]
&\FO^\pm\bigl(\eO^{\iO\,\bang{\a,x}}\,\eta\bigr)=(2\pi)^{m/2}\,\d[\pm\a]~,
\quad \a\in\X^*~.
\end{align*}

The next examples are in \hbox{$\X=\RR=\X^*$}.
We'll deal with the Heaviside step function $\he$,
the sign function $\sgn$
and the characteristic function $\chi_{[a,b]}$ of a real interval $[a,b]$\,,
which are related by the identities
(we are not interesed in values at single points)
\begin{align*}
&\he(\pm x)=\oh\,(1\pm\sgn(x))~,\qquad
\sgn(x)=\he(x)-\he(-x)~,
\\[6pt]
&\chi_{[a,b]}(x)=\he(x-a)-\he(x-b)
=\oh\,\bigl(\sgn(x-a)-\sgn(x-b)\bigr)~.
\end{align*}

We have
\begin{align*}
&\FO^\pm\bigl(\pv(1/(x-a))\bigr)(y)
=\mp\iO\,\sqrt{\tfrac\pi2}\,\eO^{\mp\iO ay}\,\sgn(y)~,\qquad
(\FO^\pm\sgn)(y)=\mp\iO\,\sqrt{\tfrac2\pi}\,\pv(1/y)~,
\\[6pt]
&(\FO^\pm\he)(y)=
\tfrac1{\sqrt{2\pi}}\,\bigl(\mp\iO\,\pv(1/y)+\pi\,\d(y)\bigr)~,\qquad
(\FO^\pm\hat\chi_{[a,b]})(y)=
\tfrac{\mp\iO}{\sqrt{2\pi}}\,(\eO^{\mp\iO by}-\eO^{\mp\iO ay})\pv(1/y)~,
\\[6pt]
&\FO^\pm\bigl(\he(ax)\cdot\exp(\iO ax)\bigr)(y)=
\tfrac1{\sqrt{2\pi}}\,\bigl(-\iO\bigl(\pv(1/(\pm y-a))+\pi\,\d[\pm y-a]\bigr)~,
\quad a\in\RR{\setminus}\{0\}~,
\\[6pt]
&\FO^\pm\bigl(\he(ax)\cdot\exp(-\iO ax)\bigr)(y)=
\tfrac1{\sqrt{2\pi}}\,\bigl(-\iO\bigl(\pv(1/(\pm y+a))+\pi\,\d[\pm y+a]\bigr)~,
\quad a\in\RR{\setminus}\{0\}~.
\end{align*}

Somewhat related to the two last formulas we also have
the physicist's integral representation of the Heaviside function:
for any $a\in\RR$ we have
\begin{align*}
&\pm2\pi\iO\,\he(\pm y)\,\eO^{\iO a y}
=\lim_{\e\to0^+}\int\limits_{-\infty}^{+\infty}
\frac{\eO^{\iO yx}}{x-a\mp\iO\e}\,\dO x
=\sqrt{2\pi}\,\lim_{\e\to0^+}\FO^-\bigl(\frac1{x-a\mp\iO\e}\bigr)(y)~,
\displaybreak[2]\\[6pt]
&\pm2\pi\iO\,\he(\mp y)\,\eO^{-\iO a y}
=\lim_{\e\to0^+}\int\limits_{-\infty}^{+\infty}
\frac{\eO^{-\iO yx}}{x-a\mp\iO\e}\,\dO x
=\sqrt{2\pi}\,\lim_{\e\to0^+}\FO^+\bigl(\frac1{x-a\mp\iO\e}\bigr)( y)~.
\end{align*}

\subsection{Elementary solutions of field equations}
\label{ss:Elementary solutions of field equations}

For simplicity we assume a fixed volume form on the base manifold $\M$.
Differential operators $D$ and $D'$, respectively acting on smooth sections
of the vector bundles \hbox{$\V\onto\M$} and \hbox{$\V^*\onto\M$},
are said to be mutually \emph{distributional adjoint} operators if
\hbox{$\bbang{D'\s,s}=\bbang{\s,Ds}$}
for any sections $s$ and $\s$ whenever the integrals are finite
(integration is performed via the fixed volume form).

In practice we deal with
\emph{polynomial derivation operators of order} \hbox{$k\in\NN$}\,,
namely with coordinate expressions of the type
$$(Ds)^i=\sum_{|\sA|\leq k} c_\sA{}^i{}_j\,\de^\sA s^j~,\qquad
(D'\s)_j=\sum_{|\sA|\leq k} c'_{\sA j}{}^i\,\de^\sA\s_i~,$$
where \hbox{${\scriptstyle A}=\{{\scriptstyle A}_1,..,{\scriptstyle A}_m\}$}
denotes multi-indices of order
\hbox{$|{\scriptstyle A}|:=\sum_a{\scriptstyle A}_a$}\,,
\hbox{$c_\sA{}^i{}_j\,,\,c'_{\sA j}{}^i:\M\to\CC$} are smooth functions, and
\hbox{$\de^\sA\equiv\de^{|\sA|}/(\de x^1)^{\sA_1}..(\de x^m)^{\sA_m}$}\,.
The natural extensions of these operators to generalized sections
maintain the same coordinate expressions.
The relation between the functions $c_\sA{}^i{}_j$
and the functions $c'_{\sA j}{}^i$ can be found by expressing
the condition \hbox{$\bbang{D'\s,s}=\bbang{\s,Ds}$} in integral form
and applying integration by parts to the first-hand side
(boundary terms disappear for test sections).
In particular,
if the coefficients $c_\sA{}^i{}_j$ are constant then we have
$$c'_{\sA j}{}^i=(-1)^{|\sA|}\,c_\sA{}^i{}_j~.$$

A couple $(D,D')$ of mutually distributional adjoint operators
also yields operators
$$D\!_\one\equiv D'\tn\id~,\quad D\!_\two\equiv \id\tn D~,$$
acting on the space of kernels (\sref{ss:Tensor products})
$$\DC\bigl(\M\times\M,\V^*\ten{\M\times\M}\V\bigr)=
\DC(\M,\V^*)\tnb\DC(\M,\V)~.$$
If \hbox{$K=K_i{}^j\,\dO y_\one^i\tn\de y_{\two j}$}
is a kernel's coordinate expression then
$$D\!_\one K=c'_{\sA i}{}^h\,\de_{\!\one}^\sA K_h{}^j\,
\dO y_\one^i\tn\de y_{\two j}~,\qquad
D\!_\two K=c_\sA{}^j{}_h\,\de_{\!\two}^\sA K_i{}^h\,
\dO y_\one^i\tn\de y_{\two j}~.$$

Let $K$ be a kernel.
For any two test sections
\hbox{$s:\M\to\V$}, \hbox{$\s:\M\to\V^*$}.
we have the generalized sections
$$K\pint\s:\M\gto\V^*~,\qquad s\pint K:\M\gto\V~,$$
and simple algebraic arguments then yield
\begin{align*}
& D'(K\pint\s)=(D\!_\one K)\pint\s~,
&& K\pint(D'\s)=(D\!_\two K)\pint\s~,
\\[6pt]
&D(s\pint K)=s\pint(D\!_\two K)~,
&& (Ds)\pint K=s\pint(D\!_\one K)~.
\end{align*}

A \emph{left} (resp.\ \emph{right}) \emph{elementary kernel} for $D'$
is defined to be a kernel $E$ such that
$$E\pint D'\s=\s~~(\text{resp. }D'(E\pint\s)=\s)~,\quad
\forall\:\s\in\DCo(\M,\V^*)~.$$
Left and right elementary kernels for $D$ are similarly defined
in the transpose kernel space.

\begin{proposition}~\par\noindent
$a)$~If $E$ is a left elementary kernel for $D'$
and \hbox{$\s\in\DCo(\M,\V^*)$}
then there exists \emph{at most} one solution \hbox{$\xi:\M\gto\V^*$}
to the differential equation \hbox{$D'\xi=\s$}\,;
if it exists then \hbox{$\xi=E\pint\s$}\,.
\smallbreak\noindent
$b)$~If $E$ is a right elementary kernel for $D'$
and \hbox{$\s\in\DCo(\M,\V^*)$}
then there exists \emph{at least} one solution \hbox{$\xi:\M\gto\V^*$}
to the differential equation \hbox{$D'\xi=\s$}\,,
given by \hbox{$\xi=E\pint\s$}\,.
\end{proposition}
\proof\par\noindent
$a)$~If \hbox{$D'\xi=\s$} and $(\xi_n)$ is a sequence in $\DCo(\M,\V^*)$
such that \hbox{$\xi_n\to\xi$}\,,
Then \hbox{$E\pint D'\xi_n=\xi_n\to\xi$}\,,
and also \hbox{$E\pint D'\xi_n\to E\pint D'\xi=E\pint\s$}\,,
so that \hbox{$E\pint\s=\xi$}.
\smallbreak\noindent
$b)$~We have \hbox{$D'(E\pint\s)=\s$}\,,
so that $E\pint\s$ is actually a solution.
\qed

So in general, if there is a left elementary kernel,
one has unicity results for the solutions of differential equations;
if there is a right elementary kernel, one has existence results.
Hence the importance of \emph{bilateral} (both left and right)
elementary kernels.
If the kernel has regularity properties
then one gets enhanced results,
holding not only for ordinary smooth sections with compact support,
but also for distributions with compact support.
Note, however, that a right elementary kernel does not solve completely
the existence problem, since in general one has to suppose that the right hand-side
of the equation has compact support.

\smallbreak
Let now $E$ be a left semi-regular kernel (\sref{ss:Tensor products}).
Then \hbox{$E\pint\s$} is a smooth ordinary section
for all \hbox{$\s\in\DCo(\M,\V^*)$},
and one gets a smooth section
\begin{align*}
&\ep:\M\to\V^*\tn\DCv(\M,\V):x\mapsto \ep_x~,\qquad
\bbang{\ep_x,\s}:=(E\pint\s)(x)~,
\\[6pt] \Rightarrow\quad
&(E\pint D'\s)(x)=\bbang{\ep_x\,,\,D'\s}=\bbang{D\ep_x\,,\,\s}~.
\end{align*}
If moreover $E$ is a left elementary kernel for $D'$ then
$$\bbang{D\ep_x\,,\,\s}=(E\pint D'\s)(x)=\s(x) \quad\Rightarrow\quad
D\ep_x=\d[x]\in\V^*_{\!\!\!x}\tn\DCv(\M,\V)~,\quad \forall\,x\in\M,$$
namely $\ep_x$ is a so-called \emph{elementary solution} for $D$,
relative to the point $x$.
Conversely, if $\ep_x$ is an elementary solution relatively for each
\hbox{$x\in\M$}, then $E$ is a left elementary kernel.

\remark~Generally speaking,
solutions of the homogeneous field equation \hbox{$D\phi=0$}
represent `free propagating' fields,
while a possible non-zero right-hand side represents a `source'.
Thus an elementary solution represents a field
in presence of an `elementary' (point-like) source.
In fundamental field theory there is no fixed source;
one rather has interactions between different fields,
each one acting as a source for the other.

\section{Special generalized densities on Minkowski spacetime}
\label{s:Special generalized densities on Minkowski spacetime}

Propagators and their applications to quantum field theories
are most naturally described in flat Minkowski spacetime
with a fixed inertial observer.
Though various types of extensions and generalizations can be devised,
here we'll work in that basic context.
So $\M$ is assumed to be an affine 4-dimensional space $\M$
whose associated vector space $\DO\M$ of ``free vectors'' is endowed
with an $\LL^2$-scaled (\sref{ss:Unit spaces}) Lorentz metric $g$\,.
We have
$$\TO\M=\M\times\DO\M~,\quad \TS\M=\M\times\DO^*\!\M~,$$
and denote future and past mass shells as
$$\ost\K{}_{\!m}^\pm\subset\DO^*\!\M~,\quad m\in\{0\}\cup\LL^{-1}.$$
Moreover we have the orthogonal splittings
$$\DO\M=\DO\M^\spar\oplus\DO\M^\sbot~,\qquad
\DO^*\!\M=(\DO^*\!\M)^\spar\oplus(\DO^*\!\M)^\sbot~,$$
into timelike and spacelike subspaces
(\ie\ parallel and orthogonal to the observer).

We'll use orthonormal linear coordinates
\hbox{$(x^\l)\equiv(x^0,x_\sbo)\equiv(x^0,x^1,x^2,x^3)$}
adapted to the above decomposition of $\DO\M$,
and denote as
\hbox{$(p_\l)\equiv(p_0,p_\sbo)\equiv(p_0,p_1,p_2,p_3)$}
the dual coordinates on $\DO^*\!\M$.
Note that these are respectively $\LL$-scaled and $\LL^{-1}$-scaled coordinates.
When the context is clear we simplify our notations
by indicating  a generic ``momentum'' as \hbox{$p\in\DO^*\!\M$},
and also using that symbol as a ``dummy variable''.
The standard volume forms on $\DO\M$ and $\DO^*\!\M$ can be written as
\begin{align*}
&\eta=\dO^4x=
\dO x^0\we\dO x^1\we\dO x^2\we\dO x^3\equiv\dO x^0\we\dO^3x_\sbo
\in\LL^4\tn\weu4\DO^*\!\M~,
\\[6pt]
&\ost\eta=\dO^4 p=
\dO p_0\we\dO p_1\we\dO p_2\we\dO p_3\equiv\dO p_0\we\dO^3 p_\sbo
\in\LL^{-4}\tn\weu4\DO\M~.
\end{align*}
We'll also use associated time-spherical coordinates
$\bigl(t,r,\th,\phi\bigr)$ on $\DO\M$ and
$\bigl(\t,\r,\vth,\vph\bigr)$ on $\DO^*\!\M$,
where \hbox{$\t\equiv p_0$} and \hbox{$t\equiv x^0$}, whence also
$$\dO^4x=r^2\,\sin\th\,\dO t\we\dO r\we\dO\th\we\dO\phi~,\quad
\dO^4p=\r^2\,\sin\vth\,\dO\t\we\dO\r\we\dO\vth\we\dO\vph~.$$
We'll also deal with the quadratic function
$$\ul g:\DO^*\!\M\to\RR\tn\LL^{-2}:
p\mapsto g^\#(p,p)\equiv p^2=p_0^2-|p_\sbo|^2$$
associated with the Lorentz metric $g^\#$ of $\DO^*\!\M$.

We'll use ``natural units'' (\sref{ss:Unit spaces}),
so that a mass is an element \hbox{$m\in\LL^{-1}$}.

\subsection{Mass-shell Leray densities} 
\label{ss:Mass-shell Leray densities}

For any given mass $m$ we use the shorthand
$$\Eo_m(p)\equiv(m^2+|p_\sbo|^2)^{1/2}\in\LL^{-1}~,$$
so that an element \hbox{$p\in\smash{\ost\K{}_{\!m}^\pm}\subset\DO^*\!\M$}
is characterized by
$$p_0=\pm\Eo_m(p)\equiv\pm\Eo_m(p_\sbo)~.$$
The Leray forms of the functions \hbox{$p\mapsto p_0\,{\mp}\,\Eo_m(p)$}
are $\LL^{-3}$-scaled densities on $\smash{\ost\K{}_{\!m}^\pm}$\,.
Seen as generalized densities on $\DO^*\!\M$,
with supports in $\smash{\ost\K{}_{\!m}^\pm}$\,, they can be written as
$$\om[p_0\,{\mp}\,\Eo_m]\equiv\d(p_0\,{\mp}\,\Eo_m)\,\dO^4p~.$$

Next we introduce the $\LL^{-2}$-scaled Leray densities
(\sref{ss:Delta-type densities})
\begin{align*}
&\om^{\pm}_m\equiv\he(\pm p_0)\,\om[\ul g\,{-}\,m^2]=
\frac1{2\,p_0}\,\om[p_0\,{\mp}\,\Eo_m]=
\pm\frac1{2\,\Eo_m}\,\om[p_0\,{\mp}\,\Eo_m]~,
\\[8pt]
&\ep^\pm_m\equiv\frac1{2\,\pi}\,\om^\pm_m=
\frac1{4\,\pi\,p_0}\,\om[p_0\,{\mp}\,\Eo_m]=
\pm\frac1{4\,\pi\,\Eo_m}\,\om[ p_0\,{\mp}\,\Eo_m]=
\frac1{4\,\pi\,p_0}\,\d( p_0\,{\mp}\,\Eo_m)\,\dO^4p~,
\\[6pt]
&\om[\ul g\,{-}\,m^2]=\om^+_m+\om^-_m=2\,\pi\,(\ep^+_m+\ep^-_m)~.
\end{align*}
In particular if $u$ is a test function on $\DO^*\!\M$ then
$$\bbang{\om[p_0\,{\mp}\,\Eo_m],u}=
\int u(\pm\Eo_m,p_\sbo)\,\dO^3p_\sbo~,\qquad
\bbang{\om^\pm_m\,,\,u}=
\pm\int\frac{\dO^3p_\sbo}{2\,\Eo_m}\,u(\pm\Eo_m\,,p_\sbo)~.$$

\remark~While the Leray form $\om[\ul g\,{-}\,m^2]$ is geometrically well-defined,
the other generalized densities introduced above
are observer-dependent.\smallbreak

The relations between $\om[\ul g\,{-}\,m^2]$
and $\om[p_0\,{\mp}\,\Eo_m]$ can be recovered by observing that
$$\dO(p_0\mp\Eo_m)\we\dO^3p_\sbo=\dO^4\,p~,$$
and that since \hbox{$p_0=\pm\Eo_m$} on $\smash{\ost\K{}_{\!m}^\pm}$\,,
there we get
$$\dO(g^\#(p,p)\,{-}\,m^2)\we
\bigl(\pm\tfrac1{\Eo_m}\,\dO^3p_\sbo\bigr)=
\pm\tfrac2{\Eo_m}\,g^{\l\m}\,p_\l\,\dO p_\m\we\dO^3p_\sbo=
\pm\tfrac{2\,p_0}{\Eo_m}\,\dO^4p=2\,\dO^4p~.$$
This also implies the coordinate expressions
$$\om[p_0\,{\mp}\,\Eo_m]=\dO^3p_\sbo~,$$
which must be interpreted as follows:
the objects $\om[p_0\,{\mp}\,\Eo_m]$\,,
seen as 3-forms on $\smash{\ost\K{}_{\!m}^\pm}$\,,
can be identified with the spatial volume form $\dO^3p_\sbo$
under the diffeomorphisms
\hbox{$\smash{\ost\K{}_{\!m}^\pm}\to(\DO^*\!\M)^\sbot$}
given by the orthogonal projection associated with the chosen observer.

It is not difficult to show, for example by formal integration,
that changing the argument's sign transforms $\om^\pm_m$
into minus one another, namely
$$\om^\pm_m\comp(-\id)=-\om^\mp_m~,\qquad \ep^\pm_m\comp(-\id)=-\ep^\mp_m~.$$

\remark~Usually, in the standard literature, the above generalized densities
are not introduced explicitely,
but rather one deals with ``delta-functions'',
that is generalized functions expressed in terms of the Dirac delta.
In the expression
\hbox{$\om[p_0\,{\mp}\,\Eo_m]\equiv \d(p_0\,{\mp}\,\Eo_m)\,\dO^4p$}
the meaning of $\d(p_0\,{\mp}\,\Eo_m)$ is essentially the traditional one,
but note that this is an $\LL$-scaled generalized function
(the relation between functions and densities on $\DO^*\!\M$ is determined
by the scaled volume form $\ost{\eta}=\dO^4p$).
On the other hand, the generalized function usually indicated as
$\d(p^2\,{-}\,m^2)$ deserves some comment.
We start from the observation that
if $a$ is a non-zero scalar constant and \hbox{$b\in\RR\tn\LL^{-1}$}\,,
then\footnote{
This result can be immediately recovered by making the variable change
\hbox{$\t'=a\,\t$} in the integral
\hbox{$\int_{-\infty}^{+\infty}\dO t\,\d(a\,(\t-b))\,u(\t)$}\,,
and considering the cases \hbox{$a\lessgtr0$} separately.} 
$$\d(a\,(\t-b))=\tfrac1{|a|}\,\d(\t-b)=\tfrac{\sgn(a)}{a}\,\d(\t-b)~.$$
Next, writing \hbox{$p^2\,{-}\,m^2=\t^2-\Eo_m^2$},
for any suitable test function $u$ and fixed $p_\sbo$ we find
$$\int\dO\t\,\he(\pm\t)\,\d\bigl(\t^2-\Eo_m^2\bigr)\,u(\t)=
\int\dO\t\,\he(\t)\,\d\bigl((\t-\Eo_m)\,(\t+\Eo_m)\bigr)\,u(\t)=
\frac1{2\,\Eo_m}\,u(\Eo_m)~.$$
Accordingly we write
\begin{align*}
&\he(\pm\t)\,\d\bigl(\t^2-\Eo_m^2\bigr)\,\dO^4p=
\frac1{2\,\Eo_m}\,\om[p_0\,{\mp}\,\Eo_m]=\pm\om^\pm_m
\\[6pt] \Rightarrow\quad
&\om[\ul g\,{-}\,m^2]=\he(\t)\,\om^+_m+\he(-\t)\,\om^-_m=
\\[6pt] &\phantom{\om[\ul g\,{-}\,m^2]}
=\bigl(\he(\t)-\he(-\t)\bigr)\,\d(\t^2-\Eo_m^2)\,\dO^4p=
\sgn(\t)\,\d(\t^2-\Eo_m^2)\,\dO^4p~.
\end{align*}

\subsection{Fourier transforms of mass-shell Leray densities}
\label{ss:Fourier transforms of mass-shell Leray densities}

The special generalized densities introduced
in~\sref{ss:Mass-shell Leray densities} are tempered,
namely they are scaled elements in $\LL^{-r}\tn\SCv(\DO^*\!\M)$
where $r$ is either 2 or 3.
Taking the orthogonal splitting of $\DO^*\!\M$
into account we obtain
$$\SCv(\DO^*\!\M)\cong\SCv((\DO^*\!\M)^\spar)\tnb\SCv((\DO^*\!\M)^\sbot)~,$$
so that we have the partial (``spatial'' and ``temporal'') Fourier transforms
\begin{align*}
&\FO^\pm_{\!\!\sbot}:\SCv(\DO^*\!\M)\to
\SCv((\DO^*\!\M)^\spar)\tnb\SCv((\DO\M)^\sbot)~,
\\[6pt]
&\FO^\pm_{\!\!\spar}:\SCv(\DO^*\!\M)\to
\SCv((\DO\M)^\spar)\tnb\SCv((\DO^*\!\M)^\sbot)~.
\end{align*}
Note that Fourier transforms of distributions on $\DO^*\!\M$ are done
via the $\LL^4$-scaled volume form \hbox{$\eta=\dO^4x$}\,,
which affects the scaling of transformed distributions accordingly.
Similarly, spatial transforms are done via the
$\LL^3$-scaled volume form \hbox{$\eta_\sbo=\dO^3x_\sbo$}\,.

\begin{lemma}\label{lemma:Fsbotepsilon}
$$\FO^+_{\!\!\sbot}\ep^\pm_m=
\FO^-_{\!\!\sbot}\ep^\pm_m=\frac{\pm\he(\pm\t-m)}{(2\pi)^{3/2}}\,
r\,\sin\th\,\sin(r\,\sqrt{\t^2-m^2})\,\dO\t\we\dO r\we\dth\we\dph~.$$
\end{lemma}\proof
If $v$ is a test function on \hbox{$(\DO^*\!\M)^\spar\times(\DO\M)^\sbot$} then
$$\FO^+_{\!\!\sbot}v(p_0\,,p_\sbo)=\tfrac{1}{(2\pi)^{3/2}}\,\int\dO^3x_\sbo\,
\eO^{-\iO\,p_\sbo x_\sbo}\,v(p_0\,,x_\sbo)~,$$
where \hbox{$\FO_{\!\!\sbot}\equiv\FO^+_{\!\!\sbot}$}\,, whence
\begin{align*}
&4\pi\,(2\pi)^{3/2}\,\bbang{\FO\!\!_\sbot\ep^\pm_m\,,\,v}=
4\pi\,(2\pi)^{3/2}\,\bbang{\ep^\pm_m\,,\,\FO\!\!_\sbot v}=
(2\pi)^{3/2}\int\frac{\dO^3p_\sbo}{\pm\Eo_m}\,\FO\!\!_\sbot v(\pm\Eo_m\,,p_\sbo)=
\\[6pt]
&\qquad
=\int\frac{\dO^3p_\sbo}{\pm\Eo_m}
\bigl(\int\dO^3x_\sbo\,\eO^{-\iO\,p_\sbo x_\sbo}\,v(\pm\Eo_m\,,x_\sbo)\bigr)=
\\[6pt]
&\qquad=
\int\dO^3x_\sbo\int\sin\vth\,\dvth\,\dvph\int_0^\infty
\frac{\r^2\,\dO\r}{\pm\t}\,\eO^{-\iO\,\r\,r\,\ff}\,v(\pm\t,x_\sbo)=
\\[6pt]
&\qquad=
\pm\int\dO^3x_\sbo\int\dvth\,\dvph\int_m^\infty\dO\t\,
\sqrt{\t^2-m^2}\,\sin\vth\,\eO^{-\iO\,\sqrt{\t^2-m^2}\,r\,\ff}\,v(\pm\t,x_\sbo)~,
\end{align*}
where we did the variable change \hbox{$\t=\Eo_m\equiv\sqrt{m^2+\r^2}$}
and used the shorthand
$$\ff\equiv\sin\th\,\sin\vth\,\cos(\phi-\vph)+\cos\th\,\cos\vth \quad\Rightarrow\quad
p_\sbo x_\sbo\equiv\bang{p_\sbo,x_\sbo}=\r\,r\,\ff~.$$
Hence we find
$$\FO_{\!\!\sbot}\ep^\pm_m=\pm\frac{\he(\pm\t-m)\,
\sqrt{\t^2-m^2}}{4\pi\,(2\pi)^{3/2}}\,
\Bigl(\int_0^\pi\!\!\dvth\int_0^{2\pi}\!\!\dvph\,\sin\vth\,
\eO^{-\iO\,\sqrt{\t^2-m^2}\,|x_\sbo|\,\ff}\Bigr)\,\dO\t\we\dO^3x_\sbo~,$$

Now the above integral apparently depends on the spherical angles
$\th$ and $\phi$ on $\DO\M$ via $\ff$\,.
However $\ep^\pm_m$ are radial,
thus $\FO_{\!\!\sbot}\ep^\pm_m$ have the same property
and we can calculate the integral for any fixed value of $\th$ and $\phi$\,.
Choosing \hbox{$\th=0$} we get \hbox{$\ff=\cos\vth$}\,,
and doing the variable change \hbox{$s=-\cos\vth$} we get
\begin{align*}
&\int_0^{2\pi}\dvph\int_0^\pi\dvth\,\sin\vth\,
\eO^{-\iO\,\sqrt{\t^2-m^2}\,|x_\sbo|\,\cos\vth}=
2\,\pi \int_{-1}^{+1}\dO s\,\eO^{\iO\,\sqrt{\t^2-m^2}\,|x_\sbo|\,s} =
\\[6pt]
&\qquad=
4\,\pi\,\frac{\sin(\sqrt{\t^2-m^2}\,|x_\sbo|)}{\sqrt{\t^2-m^2}\,|x_\sbo|}~,
\end{align*}
whence we find the stated expression for
\hbox{$\FO_{\!\!\sbot}\ep^\pm_m\equiv\FO^+_{\!\!\sbot}\ep^\pm_m$}\,.
The calculations for $\FO^-_{\!\!\sbot}\ep^\pm_m$ are similar,
and we get the same final result.
\qed

\bigbreak
In the sequel we'll be involved with the generalized functions
$$\digamma^\pm_{\!\!m}(t,r)\equiv
\int\limits_{-\infty}^{+\infty}\!\!\dO\t\,
\he(\t-m)\,\sin(r\,\sqrt{\t^2-m^2})\,\eO^{\mp\iO\,\t\,t}~,$$
which are well-defined as the Fourier transform and anti-transform
of a bounded function.
Moreover we'll use the shorthand
$$\Sigma(r,\th,\phi)\equiv\frac1{(2\pi)^{3/2}}\,r\,\sin\th\,\dO r\we\dth\we\dph
\equiv\frac{\dO^3x_\sbo}{(2\pi)^{3/2}\,|x_\sbo|}~.$$

\begin{proposition}\label{proposition:Fepsilon}
We have
\begin{align*}
&\FO^+\!\ep^+_m=\tfrac{1}{(2\pi)^2}\,\digamma^+_{\!\!m}\,\dO t\we\Sigma~,
&&\FO^+\!\ep^-_m=-\tfrac{1}{(2\pi)^2}\,\digamma^-_{\!\!m}\,\dO t\we\Sigma~,
\\[6pt]
&\FO^-\!\ep^+_m=\tfrac{1}{(2\pi)^2}\,\digamma^-_{\!\!m}\,\dO t\we\Sigma~,
&&\FO^-\!\ep^-_m=-\tfrac{1}{(2\pi)^2}\,\digamma^+_{\!\!m}\,\dO t\we\Sigma~.
\end{align*}
\end{proposition}\proof
If $w$ is a test density on \hbox{$(\DO\M)^\spar\times(\DO\M)^\sbot)$}
then using lemma~\ref{lemma:Fsbotepsilon} we get
\begin{align*}
&\qquad\bbang{\FO\ep^\pm_m\,,\,w}=\bbang{\FO\!_\spar\FO\!\!_\sbot\ep^\pm_m\,,\,w}
=\bbang{\FO\!\!_\sbot\ep^\pm_m\,,\,\FO\!_\spar w}=
\\[6pt]
&=\frac{\pm1}{(2\pi)^2}
\int\limits_{-\infty}^{+\infty}\!\!\dO\t\,\he(\pm\t-m)
\int\limits_0^\infty\!\!r\,\dO r
\int\limits_0^\pi\!\!\sin\th\,\dth\int\limits_0^{2\pi}\!\!\dph\,
\sin(r\,\sqrt{\t^2-m^2})
\int\limits_{-\infty}^{+\infty}\!\!\dO t\,\eO^{-\iO\,\t\,t}\,
w(t,r,\th,\phi)=
\\[6pt]
&=\frac{1}{(2\pi)^2}
\int\limits_0^\infty\!\!r\,\dO r\!
\int\limits_0^\pi\!\!\sin\th\,\dth\! \int\limits_0^{2\pi}\!\!\dph\!
\int\limits_{-\infty}^{+\infty}\!\!\dO\t\,\bigl(\pm\he(\pm\t-m)\bigr)
\sin(r\,\sqrt{\t^2-m^2})\!
\int\limits_{-\infty}^{+\infty}\dO t\,\eO^{-\iO\,\t\,t}\,
w(t,r,\th,\phi)\,,
\end{align*}
whence the stated expression for $\FO^+\!\ep^+_m$ follows.
The other expressions follow from suitable variable changes.
\qed

\begin{corollary}\label{corollary:Fepmalgebraicrels}
We have the identities
$$\FO^+\!\ep^+_m+\FO^-\!\ep^-_m=0~,\qquad\FO^+\!\ep^-_m+\FO^-\!\ep^+_m=0~.$$
Moreover at \hbox{$t=0$} we have
$$\FO^+\!\ep^+_m{\bigl|}_{t=0}=
\FO^-\!\ep^+_m{\bigl|}_{t=0}=
-\FO^+\!\ep^-_m{\bigl|}_{t=0}=
-\FO^-\!\ep^-_m{\bigl|}_{t=0}~.$$
\end{corollary}

We now introduce a further set of generalized densities
related to mass-shells, namely
\begin{align*}
&\opp{m}{+}{+}\equiv
\lim_{\e\to0^+}\frac{\dO^4p}{\Eo_m\,(p_0+\Eo_m+\iO\e)}~,
&&\opp{m}{+}{-}\equiv
\lim_{\e\to0^+}\frac{\dO^4p}{\Eo_m\,(p_0+\Eo_m-\iO\e)}~,
\\[6pt]
&\opp{m}{-}{+}\equiv
\lim_{\e\to0^+}\frac{\dO^4p}{\Eo_m\,(p_0-\Eo_m+\iO\e)}~,
&&\opp{m}{-}{-}\equiv
\lim_{\e\to0^+}\frac{\dO^4p}{\Eo_m\,(p_0-\Eo_m-\iO\e)}~.
\end{align*}

Denoting Fourier transforms and anti-trasforms
respectively by hats and an overchecks we then find
\begin{lemma}\label{lemma:halfpropagators}
We have
\begin{align*} 
\fpp{m}{+}{+}(t,r,\th,\phi)&=
-2\,\iO\,\he(t)\,\digamma^-_{\!\!m}(t,r)\,
\dO t\we\Sigma(r,\th,\phi)~,
\displaybreak[2]\\[6pt] 
\vpp{m}{+}{+}(t,r,\th,\phi)&=
-2\,\iO\,\he(-t)\,\digamma^+_{\!\!m}(t,r)\,
\dO t\we\Sigma(r,\th,\phi)
\displaybreak[2]\\[6pt] 
\fpp{m}{-}{+}(t,r,\th,\phi)&=
-2\,\iO\,\he(t)\,\digamma^+_{\!\!m}(t,r)\,
\dO t\we\Sigma(r,\th,\phi)~,
\displaybreak[2]\\[6pt] 
\vpp{m}{-}{+}(t,r,\th,\phi)&=
-2\,\iO\,\he(-t)\,\digamma^-_{\!\!m}(t,r)\,
\dO t\we\Sigma(r,\th,\phi)~,
\displaybreak[2]\\[6pt] 
\fpp{m}{+}{-}(t,r,\th,\phi)&=
2\,\iO\,\he(-t)\,\digamma^-_{\!\!m}(t,r)\,
\dO t\we\Sigma(r,\th,\phi)~,
\displaybreak[2]\\[6pt] 
\vpp{m}{+}{-}(t,r,\th,\phi)&=
2\,\iO\,\he(t)\,\digamma^+_{\!\!m}(t,r)\,
\dO t\we\Sigma(r,\th,\phi)~,
\displaybreak[2]\\[6pt] 
\fpp{m}{-}{-}(t,r,\th,\phi)&=
2\,\iO\,\he(-t)\,\digamma^+_{\!\!m}(t,r)\,
\dO t\we\Sigma(r,\th,\phi)~,
\displaybreak[2]\\[6pt] 
\vpp{m}{-}{-}(t,r,\th,\phi)&=
2\,\iO\,\he(t)\,\digamma^-_{\!\!m}(t,r)\,
\dO t\we\Sigma(r,\th,\phi)~.
\end{align*}
\end{lemma}\proof
We carry out the explicit calculations for the first identity;
the others follow similarly.
We use the same argument about radial symmetry as in the proof
of lemma~\ref{lemma:Fsbotepsilon},
and the integral representations of the Heaviside distribution
(\sref{ss:Fourier transforms}).
Doing at at the proper passage the variable change
\hbox{$\s=\Eo_m=\sqrt{m^2+\r^2}$},
for any test function \hbox{$u\in\SCo(\DO\M)$} we obtain
\begin{align*}
& \bbang{\fpp{m}{+}{+}\;,u}=\bbang{\opp{m}{+}{+}\;,\hat u}=
\\[8pt]
& =\frac1{(2\pi)^2}
\int\limits_{-\infty}^{+\infty}\!\!\! \dO t
\int\limits_0^{+\infty}\!\!\! r^2\,\dO r
\int\limits_0^\pi\!\! \sin\th\,\dth
\int\limits_0^{2\pi}\!\! \dph
\int\limits_{-\infty}^{+\infty}\!\!\! \dO\t
\int\limits_0^{+\infty}\!\!\! \r^2\,\drh
\int\limits_0^\pi\!\! \sin\vth\,\dvth
\int\limits_0^{2\pi}\!\! \dvph
\frac{\eO^{-\iO\t t}\,\eO^{-\iO\r r \ff}}{\Eo_m\,(\t+\Eo_m+\iO\e)}\,
u(t, r,\th,\phi)
\displaybreak[1]\\[8pt]
&\xrightarrow{\e\to0^+}\,\frac1{(2\pi)^2}
\int\limits_{-\infty}^{+\infty}\!\!\! \dO t
\int\limits_0^{+\infty}\!\!\!  r^2\,\dO r
\int\limits_0^\pi\!\! \sin\th\,\dth
\int\limits_0^{2\pi}\!\! \dph
\int\limits_0^{+\infty}\!\!\!\frac{\r^2}{\Eo_m}\,\drh\,
2\pi\iO\,\eO^{+\iO\,\Eo_m\,t}\,(-\he(t))
\frac{4\pi}{\r r}\sin(\r r)\,u(t,r,\th,\phi)=
\displaybreak[2]\\[8pt]
& =-2\,\iO
\int\limits_{-\infty}^{+\infty}\!\!\! \he(t)\,\dO t
\int\limits_0^{+\infty}\!\!\!  r\,\dO r
\int\limits_0^\pi\!\! \sin\th\,\dth
\int\limits_0^{2\pi}\!\! \dph\,\,u(t, r,\th,\phi)
\int\limits_0^{+\infty}\!\!\drh\,\frac\r{\Eo_m}\,
\sin( r\r)\,\eO^{+\iO\,\Eo_m\,t}=
\displaybreak[2]\\[8pt]
& =-2\,\iO
\int\limits_{-\infty}^{+\infty}\!\!\! \he(t)\,\dO t
\int\limits_0^{+\infty}\!\!\! r\,\dO r
\int\limits_0^\pi\!\! \sin\th\,\dth
\int\limits_0^{2\pi}\!\! \dph\,\,u(t,r,\th,\phi)
\int\limits_m^{+\infty}\!\!\dO\s\,
\sin( r\,\sqrt{\s^2-m^2})\,\eO^{+\iO\,\s\,t}=
\displaybreak[2]\\[8pt]
& =-2\,\iO
\int\limits_{-\infty}^{+\infty}\!\!\! \he(t)\,\dO t
\int\limits_0^{+\infty}\!\!\! r\,\dO r
\int\limits_0^\pi\!\! \sin\th\,\dth
\int\limits_0^{2\pi}\!\! \dph\,
\digamma^-_{\!\!m}(t,r)\,u(t,r,\th,\phi)~.
\end{align*}
\qed

\begin{proposition}\label{proposition:epsilontoo}
\begin{align*}
&2\,(2\pi)^2\,\iO\,\ep^+_m=\opp{m}{-}{-}-\opp{m}{-}{+}~,
\\[6pt]
&2\,(2\pi)^2\,\iO\,\ep^-_m=\opp{m}{+}{+}-\opp{m}{+}{-}~.
\end{align*}
\end{proposition}\proof
Both identities can be obtained by comparing either the Fourier transforms
or the Fourier anti-trasforms of their two sides,
taking  proposition~\ref{proposition:Fepsilon}
and lemma~\ref{lemma:halfpropagators} into account.\qed

\remark~In physics texts, approximate expressions like
$$\frac{1}{\Eo_m\,(p_0-\Eo_m+\iO\e)}-\frac{1}{\Eo_m\,(p_0+\Eo_m+\iO\e)}
~\stackrel{\phantom{{}^+}\e\to0^+}{\approx}~
\frac{1}{p^2-m^2+\iO\,\e}$$
are commonly used.
We note, however, that the right-hand side does not characterize
a unique combination of the generalized densities $\opp{m}{\pm}{\pm}$\,,
so we prefer to avoid such shortcuts in order to keep a fairly
systematic approach.

\subsection{Mass-shell related principal values}
\label{ss:Mass-shell related principal values}

Consider the scaled generalized densities
\hbox{$e^\pm_m\in\LL^{-2}\tn\SCv(\DO^*\!\M)$} defined by
\begin{align*}
&4\,\pi^2\,\bang{e^+_m\,,\,v}=\lim_{\e\to0^+}\int\!\!\dO^3p_\sbo
\Bigl(\int\limits_{-\infty}^{\Eo_m-\e}\!\!\!\!\dO p_0~
+\int\limits_{\Eo_m+\e}^{+\infty}\!\!\!\!\dO p_0\Bigr)
\frac{v(p_0,p_\sbo)}{\Eo_m\,(p_0-\Eo_m)}~,
\\[6pt]
&4\,\pi^2\,\bang{e^-_m\,,\,v}=\lim_{\e\to0^+}\int\!\!\dO^3p_\sbo
\Bigl(\int\limits_{-\infty}^{-\Eo_m-\e}\!\!\!\!\!\!\dO p_0~
+\int\limits_{-\Eo_m+\e}^{+\infty}\!\!\!\!\!\!\dO p_0\Bigr)
\frac{v(p_0,p_\sbo)}{\Eo_m\,(p_0+\Eo_m)}~.
\end{align*}
Recalling \sref{ss:Principal value densities} we then see
that $e^\pm_m$ are principal value densities:
$$4\,\pi^2\,e^\pm_m
=\pv\frac{\dO^4p}{\Eo_m\,(p_0\mp\Eo_m)}~.$$
Actually, setting \hbox{$\s\equiv p_0\mp\Eo_m$}
we obtain new coordinates $\bigl(\s,\r,\vth,\vph\bigr)$
on the future and past causal cones,
with the $m$-shells characterized by \hbox{$\s=0$}\,,
and we have
$$4\,\pi^2\,e^\pm_m=\frac1{\Eo_m}\,\pv\frac{\dO^4p}\s=
\pv\frac{\dO^4p}{\Eo_m\,\s}~.$$

Moreover we set
$$e_m\equiv e^+_m-e^-_m=
\frac{1}{2\,\pi^2}\,\pv\frac{\dO^4p}{p^2-m^2}~,$$
and note that this is a geometrically well-defined generalized density
(while $e^\pm_m$ are observer-dependent).
We have the identity \hbox{$e^\pm_m(-p)=-e^\mp_m(p)$}\,,
which also implies that \hbox{$e_m\equiv e^+_m-e^-_m$}
is symmetric in its argument, as one expects from its above expression
as the principal value of a symmetric density.

\begin{proposition}\label{proposition:etoo}
We have
\begin{align*}
&(2\pi)^2\,\hat e^\pm_m=
-\iO\,\sgn(t)\,\digamma^\pm_{\!\!m}(t,r)\,\dO t\we\Sigma(r,\th,\phi)~,
\\[10pt]
&\hat e_m=\check e_m=
\iO\,\sgn(t)\,\FO^-\!\bigl(\ep^+_m + \ep^-_m\bigr)~,
\end{align*}
whence also
$$2\,(2\pi)^2\,e^+_m=\opp{m}{-}{-}+\opp{m}{-}{+}~,\qquad
2\,(2\pi)^2\,e^-_m=\opp{m}{+}{-}+\opp{m}{+}{+}~.$$
\end{proposition}\proof
For \hbox{$u\in\SCo(\DO\M^\spar\times\DO^*\!\M^\sbot)$},
using \hbox{$\FO\bigl(\pv(1/(x-a))\bigr)(y)=
-\iO\,\sqrt{\frac\pi2}\,\eO^{-\iO ay}\,\sgn(y)$}
we get
\begin{align*}
&4\,\pi^2\,\bbang{\FO_{\!\spar}e^\pm_m\,,\,u}=
 4\,\pi^2\,\bbang{e^\pm_m\,,\,\FO_{\!\spar}u}=
\\[6pt]
&\qquad=\lim_{\e\to0^+}\int\!\!\r^2\,\sin\vth\,\dO\r\,\dvth\,\dvph\,
\Bigl(\int\limits_{-\infty}^{\pm\Eo_m-\e}\!\!\!\!\dO\t\,\,
{+}\!\!\!\int\limits_{\pm\Eo_m+\e}^{+\infty}\!\!\!\!\dO\t\Bigr)\!\!
\int\limits_{-\infty}^{+\infty}\!\!\dO t\,
\frac{\eO^{-\iO\,\t\,t}\,u(t,\r,\vth,\vph)}{\sqrt{2\pi}\,\Eo_m\,(\t\mp\Eo_m)}=
\displaybreak[2]\\[6pt]
&\qquad=\int\!\!\frac{\r^2}{\Eo_m}\,\sin\vth\,\dO\r\,\dvth\,\dvph
\int\limits_{-\infty}^{+\infty}\!\!\dO t\,u(t,\r,\vth,\vph)\,\lim_{\e\to0^+}
\Bigl(\int\limits_{-\infty}^{\pm\Eo_m-\e}\!\!\!\!\dO\t\,\,
{+}\!\!\!\int\limits_{\pm\Eo_m+\e}^{+\infty}\!\!\!\!\dO\t\Bigr)
\frac{\eO^{-\iO\,\t\,t}}{\sqrt{2\pi}\,(\t\mp\Eo_m)}=
\displaybreak[2]\\[6pt]
&\qquad=\int\!\!\frac{\r^2}{\Eo_m}\,\sin\vth\,\dO\r\,\dvth\,\dvph
\int\limits_{-\infty}^{+\infty}\!\!\dO t\,u(t,\r,\vth,\vph)\,
\int\limits_{-\infty}^{+\infty}\!\!\dO\t
\frac{\eO^{-\iO\,\t\,t}}{\sqrt{2\pi}}\,\pv\frac1{\t\mp\Eo_m}=
\displaybreak[2]\\[6pt]
&\qquad=-\iO\,\sqrt{\frac\pi2}
\int\!\!\frac{\r^2}{\Eo_m}\,\sin\vth\,\dO\r\,\dvth\,\dvph
\int\limits_{-\infty}^{+\infty}\!\!\dO t\,u(t,\r,\vth,\vph)\,
\eO^{\mp\iO\,\Eo_m t}\,\sgn(t) \quad\Rightarrow
\displaybreak[2]\\[8pt]
\Rightarrow~&\FO_{\!\!\spar}e^\pm_m=
\frac{-\iO}{2\,(2\pi)^{3/2}}\,
\sgn(t)\,\eO^{\mp\iO\,\Eo_m t}\,\frac{\r^2}{\Eo_m}\,\sin\vth\,
\dO t\we\dO\r\we\dvth\we\dvph~.
\end{align*}
Using the variable change \hbox{$\t=\Eo_m=\sqrt{m^2+\r^2}$}
and the spherical symmetry argument as in the proof
of lemma~\ref{lemma:Fsbotepsilon} we find
\begin{align*}
\FO_{\!\!\sbot}\Bigl(\frac{\eO^{\mp\iO\,\Eo_m t}}{\Eo_m}\,\dO^3p_\sbo\Bigr)&=
\frac{\dO^3x_\sbo}{(2\pi)^{3/2}}\int\r^2\,\sin\vth\,\dO\r\,\dvth\,\dvph\,
\frac{\eO^{\mp\iO\,\Eo_m t}\,\eO^{-\iO\,\r\,r\,\ff}}{\Eo_m}=
\\[6pt]
&=\sqrt{\tfrac2\pi}\,\digamma^\pm_{\!\!m}(t,r)\,\Sigma(r,\th,\phi)~.
\end{align*}
We then get our statement using
\hbox{$\FO e^\pm_m=\FO_{\!\!\sbot}\FO_{\!\!\spar}e^\pm_m$}\,,
\hbox{$e_m\equiv e^+_m-e^-_m$}
and lemma~\ref{lemma:halfpropagators}.\qed

\begin{corollary}
Summarising propositions~\ref{proposition:epsilontoo}
and~\ref{proposition:etoo} we have the relations
\begin{align*}
&2\,(2\pi)^2\,\iO\,\ep^+_m=\opp{m}{-}{-}-\opp{m}{-}{+}~,
&&2\,(2\pi)^2\,\iO\,\ep^-_m=\opp{m}{+}{+}-\opp{m}{+}{-}~,
\\[6pt]
&2\,(2\pi)^2\,e^+_m=\opp{m}{-}{-}+\opp{m}{-}{+}~,
&&2\,(2\pi)^2\,e^-_m=\opp{m}{+}{-}+\opp{m}{+}{+}~,
\end{align*}
which can be inverted as
\begin{align*}
&\opp{m}{-}{-}=(2\pi)^2\,(e^+_m+\iO\,\ep^+_m)~,
&&\opp{m}{-}{+}=(2\pi)^2\,(e^+_m-\iO\,\ep^+_m)~,
\\[6pt]
&\opp{m}{+}{-}=(2\pi)^2\,(e^-_m-\iO\,\ep^-_m)~,
&&\opp{m}{+}{+}=(2\pi)^2\,(e^-_m+\iO\,\ep^-_m)~.
\end{align*}
\end{corollary}

\subsection{Elementary solutions of the Klein-Gordon equation}
\label{ss:Elementary solutions of the Klein-Gordon equation}

As briefly discussed in the remark
concluding~\sref{ss:Elementary solutions of field equations},
if $D$ is a differential operator on $\DO\M$
then one is interested in solutions of the equation \hbox{$D\xi=\d[0]$},
called ``elementary solutions'' in relation to the
homogeneous equation \hbox{$D\xi=0$}\,,
the general idea being that the solutions of the latter are ``free fields'',
while elementary solutions are the buiding blocks of field interactions.
The main point of our discussion is now then that the problem of finding
and classifying elementary solutions can be converted,
via Fourier transform, to a problem of division (\sref{ss:Division}).

The Klein-Gordon operator,
acting on $\CC$-valued distributions in flat spacetime,
is the $\LL^{-2}$-scaled operator \hbox{$\dAl\,\,{+}\,m^2$},
where \hbox{$\dAl\,\equiv g^{\l\m}\,\de_\l\,\de_\m$}
is the \emph{d'Alembertian} (or \emph{wave operator}).
The elementary solutions of the Klein-Gordon equation are then
the generalized densities $\xi$ on $\DO\M$ fulfilling
$$(\dAl\,+m^2\,)\,\xi=\d[0]~.$$
By taking the Fourier transform of both members,
we see that $\xi$ is an elementary solution
if and only if its Fourier transform $\hat\xi$ fulfills
$$(2\pi)^2\,(-\ul g+m^2)\,\hat\xi=\ost\eta \quad\Leftrightarrow\quad
(2\pi)^2\,\hat\xi=\ost\eta/(-\ul g+m^2)~,$$
where \hbox{$\ul g(p)\equiv g^\#(p,p)\equiv p^2$}.
Also, $\xi$ is a solution of the homogeneous equation
if and only if \hbox{$(-\ul g+m^2)\,\hat\xi=0$}\,.

\begin{proposition}\label{proposition:basicKGelementarysolutions}
~\smallbreak\noindent
$(i)$~The generalized section $-\oh\,\check e_m$ is an elementary solution
of the Klein-Gordon equation;
\smallbreak\noindent
$(ii)$~the generalized sections $\check\ep^\pm_m$ are solutions
of the (homogeneous) Klein-Gordon equation.
\end{proposition}\proof~$(i)$
From the basics about principal values (\sref{ss:Principal value densities})
and division (\sref{ss:Division}) we have
$$\ost{\eta}\equiv\dO^4p=
(2\pi)^2\,(-\ul g+m^2)\,
\bigl(-\frac{1}{4\,\pi^2}\,\pv\frac{\dO^4\pp}{\ul g-m^2}\bigr)=
(2\pi)^2\,(-\ul g+m^2)\,(-\oh\,e_m)~.$$
$(ii)$~The vanishing of \hbox{$(-\ul g+m^2)\,\ep^\pm_m$}
follows from general arguments~(\sref{ss:Division}),
and can also be directly checked by
$$\ep^\pm_m=\frac{\pm1}{4\,\pi\,\Eo_m}\,\d(\pp_0-\Eo_m)\,\dO^4\pp \quad\Rightarrow\quad
(-\ul g+m^2)\,\ep^\pm_m=
\pm\frac{-\Eo_m^2+|p_\sbo|^2+m^2}{4\,\pi\,\Eo_m}\,\dO^4p=0~.$$
\qed

So we can construct elementary solutions
by adding to $-\oh\,e_m$ any linear combination of $\ep^\pm_m$
and then taking the inverse Fourier transform.\footnote{
More generally, any generalized density on mass-shell
determines a solution of the homogeneous equation.} 
We'll be particularly involved
with the combinations \hbox{$\pm\ih\,(\ep^+_m \pm \ep^-_m)$}\,.

\begin{proposition}\label{proposition:Fm_elementaryKGsolutions}
We have the elementary solutions
\begin{align*}
&-\oh\,\check e_m+\ih\,(\check\ep^+_m + \check\ep^-_m)=
\iO\,\he(-t)\,(\check\ep^+_m + \check\ep^-_m)~,
\\[6pt]
&-\oh\,\check e_m-\ih\,(\check\ep^+_m + \check\ep^-_m)=
-\iO\,\he(t)\,(\check\ep^+_m + \check\ep^-_m)~,
\\[6pt]
&-\oh\,\check e_m+\ih\,(\check\ep^+_m - \check\ep^-_m)=
\iO\,\bigl(\he(-t)\,\check\ep^+_m-\he(t)\,\check\ep^-_m\bigr)~,
\\[6pt]
&-\oh\,\check e_m-\ih\,(\check\ep^+_m - \check\ep^-_m)=
\iO\,\bigl(\he(-t)\,\check\ep^-_m-\he(t)\,\check\ep^+_m\bigr)~.
\end{align*}
\end{proposition}\proof
It follows from the above considerations and the identity
\hbox{$\check e_m=\iO\,\sgn(t)\,(\check\ep^+_m + \check\ep^-_m)$}\,,
found in proposition~\ref{proposition:etoo}.\qed

\subsection{Klein-Gordon propagators}
\label{ss:Klein-Gordon propagators}

Spacetime propagators (also called \emph{Green functions})
essentially arise as Fourier transforms and anti-transforms
of combinations of the generalized densities $e_m$ and $\ep^\pm_m$\,,
in relation to the classification of elementary solutions
of the Klein-Gordon equation,
and also appear in the evaluation of quantum field super-commutators.
These objects can be seen, via the distinguished volume form $\eta$\,,
as generalized functions on $\DO\M$.
We shall denote them by symbols of the type $\Dcal^\square_{\!m}$\,,
where the square is to be replaced by some label,
and by \hbox{$\Dcal^\square\equiv\Dcal^\square_{\!0}$}
in the massless case \hbox{$m=0$}\,.
The corresponding \emph{two-point propagators}
are essentially kernels on $\M$
(\sref{ss:Elementary solutions of field equations})
written as \hbox{$K(x,x')=\Dcal^\square(x'-x)$}\,.

In the literature, propagators are introduced
in various slightly different forms as generalized integrals,
and a precise comparison among these is not always immediate;
moreover there may be differences in basic conventions
(see also the remark ending this section).

Considering the integrals
$$\Dcal^\pm_{\!m}(t,x_\sbo)\equiv\Dcal^\pm_{\!m}(x)\equiv
\frac{\pm1}{(2\pi)^{3}}\int\frac{\dO^3p_\sbo}{2\,\Eo_m(p_\sbo)}\,
\eO^{\mp\iO\,(\Eo_m(p_\sbo)t+p_\sbo x_\sbo)}~,$$
we observe that $\Dcal^\pm_{\!m}$ are generalized functions
which can be respectively seen as the  Fourier transform and anti-transform of
the Leray density of the future mass-shell.
Indeed these apparently spatial integrals can be conveniently rewritten as
\begin{align*}
&\Dcal^\pm_{\!m}(x)=\frac{\pm1}{(2\pi)^{3}}
\int\frac{\dO^4p}{2\,\Eo_m(p_\sbo)}\,
\eO^{\mp\iO\,p\,x}\,\d\bigl(p_0-\Eo_m(p_\sbo)\bigr) \quad\Rightarrow
\\[8pt] \Rightarrow\quad
&\Dcal^\pm_{\!m}\,\eta=
\pm\FO^{\pm}\ep^+_m=\mp\FO^\mp\ep^-_m=
-\check\ep^\mp_m=+\hat\ep^\pm_m~.
\end{align*}

Moreover we introduce the combinations
$$\Dcal_{\!m}\equiv\Dcal^+_{\!m} + \Dcal^-_{\!m}~,\qquad
\Dcal^\circ_{\!m}\equiv\Dcal^+_{\!m} - \Dcal^-_{\!m}~.$$
Using definitions and remarks in~\sref{ss:Mass-shell Leray densities}
we can also write
\begin{align*}
\Dcal^\pm_{\!m}\,\eta&=
\frac{\pm1}{2\,\pi}\,\FO^{\pm}\bigl(\he(p_0)\,\d(p_0^2-\Eo_m^2)\bigr)=
\frac{\pm1}{(2\pi)^{3}}\int\!\dO^4p\,
\eO^{\mp\iO\,p\,x}\,\he(p_0)\,\d(p_0^2-\Eo_m^2)=
\\[8pt]
&=\frac{\pm1}{2\,\pi}\,\FO^{\mp}\bigl(\he(-p_0)\,\d(p_0^2-\Eo_m^2)\bigr)=
\frac{\pm1}{(2\pi)^{3}}\int\!\dO^4p\,
\eO^{\pm\iO\,p\,x}\,\he(-p_0)\,\d(p_0^2-\Eo_m^2)~,
\\[10pt]
\Dcal_{\!m}\,\eta&=\frac{1}{(2\pi)^{3}}\int\!\dO^4p\,\eO^{-\iO\,p\,x}\,
\sgn(p_0)\,\d(p_0^2-\Eo_m^2)=
\frac{\pm1}{2\pi}\,\FO^{\pm}\!\om[\ul g-m^2]=
\\[8pt]
&=\FO^+\!(\ep^+_m + \ep^-_m)=-\FO^-\!(\ep^+_m + \ep^-_m)=
\frac{1}{(2\,\pi)^2}\,
\bigl(\digamma^+_{\!\!m}-\digamma^-_{\!\!m}\bigr)\,\dO t\we\Sigma~.
\end{align*}
Similarly we find
$$\Dcal^\circ_{\!m}\,\eta=\FO^+\!(\ep^+_m - \ep^-_m)=\FO^-\!(\ep^-_m - \ep^+_m)=
\frac{1}{(2\,\pi)^2}\,
\bigl(\digamma^+_{\!\!m}+\digamma^-_{\!\!m}\bigr)\,\dO t\we\Sigma~.$$

We easily check the identities
$$\Dcal^+_{\!m}(-x)=-\Dcal^-_{\!m}(x) \quad\Rightarrow\quad 
\begin{cases}
\Dcal_{\!m}(-x)=-\Dcal_{\!m}(x)~,\\[6pt]
\Dcal^\circ_{\!m}(-x)=\Dcal^\circ_{\!m}(x)~.\end{cases}$$
Moreover from corollary~\ref{corollary:Fepmalgebraicrels},
or directly from the definition of $\digamma^\pm_{\!\!m}$\,,
we also see that at \hbox{$t=0$} we have
$$\Dcal^+_{\!m}(0,x_\sbo)=-\Dcal^-_{\!m}(0,x_\sbo) \quad\Rightarrow\quad
\begin{cases}\Dcal_{\!m}(0,x_\sbo)=0~,\\[6pt]
\Dcal^\circ_{\!m}(0,x_\sbo)=2\,\Dcal^+_{\!m}(0,x_\sbo)=
-2\,\Dcal^-_{\!m}(0,x_\sbo)~.
\end{cases}~.$$

\remark~Since the generalized densities $\ep^\pm_m$ are observer-dependent,
so are the generalized functions $\Dcal^\pm_{\!m}$\,;
but their sum $\Dcal_{\!m}$ is a geometrically well-defined object,
as it is the Fourier transform of the observer-independent Leray form
$\om[\ul g\,{-}\,m^2]$\,.
This implies that $\Dcal_{\!m}$ vanishes outside the causal cone, that is
\hbox{$g(x,x)<0~\Rightarrow~ \Dcal_{\!m}(x)=0$}\,.\smallbreak

We also consider the partial derivatives
$$\Dcal^\pm_{\!m,\l}(x)=
\frac{-\iO}{(2\pi)^{3}}\int\!\!\dO^4p\,\frac{p_\l}{2\,\Eo_m(p_\sbo)}\,
\eO^{\mp\iO\,p\,x}\,\d\bigl(p_0-\Eo_m(p_\sbo)\bigr)~,$$
and in particular the time derivatives
\begin{align*}
\Dcal^\pm_{\!m,0}(t,x_\sbo)&\equiv\Dcal^\pm_{\!m,0}(x)=
\frac{-\iO}{2\,(2\pi)^{3}}\int\!\!\dO^4p\,
\eO^{\mp\iO\,p\,x}\,\d\bigl(p_0-\Eo_m(p_\sbo)\bigr)=
\\[6pt]
&=\frac{-\iO}{2\,(2\pi)^{3}}\int\!\!\dO^3p_\sbo\,
\eO^{\mp\iO\,(\Eo_m(p_\sbo)\,t+p_\sbo x_\sbo)}~.
\end{align*}
At \hbox{$t=0$}\,, in particular, we find
$$\Dcal^\pm_{\!m,0}(0,x_\sbo)=
\frac{-\iO}{2\,(2\pi)^{3}}\int\!\!\dO^3p_\sbo\,
\eO^{\mp\iO\,p_\sbo x_\sbo}=-\ih\,\d(x_\sbo)~,$$
whence
$$\Dcal_{\!m,0}(0,x_\sbo)=-\iO\,\d(x_\sbo)~,\qquad
\Dcal_{\!m,0}^\circ(0,x_\sbo)=0~.$$

Summarizing,
the four above introduced generalized functions
are related to generalized densities
solutions of the homogeneous Klein-Gordon equation
(\sref{ss:Elementary solutions of the Klein-Gordon equation}) by
$$\Dcal^+_{\!m}\,\eta=-\check\ep^-_m~,\qquad
\Dcal^-_{\!m}\,\eta=-\check\ep^+_m~,\qquad
\Dcal_{\!m}\,\eta=-(\check\ep^+_m+\check\ep^-_m\bigr)~,\qquad
\Dcal^\circ_{\!m}\,\eta=\check\ep^+_m-\check\ep^-_m~.$$
We obtain elementary solutions
(\emph{not} all possible elementary solutions)
by adding any combination of these to
$$\iO\,\Dcal^\bullet_{\!m}\,\eta\equiv
-\oh\,\check e_m=-\ih\,\sgn(t)\,(\check\ep^+_m + \check\ep^-_m)~.$$
In particular, we obtain the elementary solutions
$$\iO\,\Dcal^\ret_{\!m}\,\eta\equiv
(\iO\,\Dcal^\bullet_{\!m}+\ih\,\Dcal_{\!m})\,\eta~,\qquad
\iO\,\Dcal^\adv_{\!m}\,\eta\equiv
(\iO\,\Dcal^\bullet_{\!m}-\ih\,\Dcal_{\!m})\,\eta~,\qquad
\iO\,\Dcal^\Fey_{\!m}\,\eta\equiv
(\iO\,\Dcal^\bullet_{\!m}+\ih\,\Dcal^\circ_{\!m})\,\eta~,$$
and the identities
\begin{align*}
&\oh\,(\Dcal^\ret_{\!m}+\Dcal^\adv_{\!m})=\Dcal^\bullet_{\!m}~,\qquad
\Dcal^\ret_{\!m}-\Dcal^\adv_{\!m}=\Dcal_{\!m}~,\\[6pt]
&\Dcal^\ret_{\!m}=\he(t)\,\Dcal_{\!m}~,\qquad
\Dcal^\adv_{\!m}=-\he(-t)\,\Dcal_{\!m}~,\qquad
\Dcal^\Fey_{\!m}=\he(t)\,\Dcal^+_{\!m}-\he(-t)\,\Dcal^-_{\!m}~.
\end{align*}
The generalized functions $\Dcal^\ret_{\!m}$\,,
$\Dcal^\adv_{\!m}$ and $\Dcal^\Fey_{\!m}$
are usually referred to as the \emph{retarded}, \emph{advanced}
and \emph{Feynman propagators}, respectively.
All of them are even.

\remark~There is no uniformity in the literature
about the precise notations and conventions related
to the family of generalized functions we denoted by the symbol $\Dcal$.
Common choices are $\Delta$\,, $D$ or $G$.
Precise definitions may differ by factors $\pm\iO$
or given in a slightly different form,
and may also depend from other conventions
(e.g.\ the spacetime metric signature).
Physics authors tend to focus on the practical formulas
they effectively need rather than on a systematic treatment.
Barut~\cite{Ba80} is one text which gives more details; there,
the symbols $\Delta^1$ and $\bar\Delta$ essentially correspond
to our $\Dcal^\circ$ and $\Dcal^\bullet$.

\vfill\newpage

\subsection{Massless case and wave equation}
\label{ss:Massless case and wave equation}
\subsubsection{Special spacetime densities in the massless case}
\label{sss:Special spacetime densities in the massless case}

When \hbox{$m=0$} then \hbox{$\Eo_m(p)=|p_\sbo|$}
and the involved shells are the future and past light cones
\hbox{$\smash{\ost\K{}_{\!0}^\pm}\subset\DO^*\!\M$}.
The definitions and results of the previous sections still hold.
In particular
\begin{align*}
&\ep^\pm_0\equiv
\frac1{4\,\pi\,p_0}\,\om[p_0\,{\mp}\,|p_\sbo|]=
\pm\frac1{4\,\pi\,|p_\sbo|}\,\om[p_0\,{\mp}\,|p_\sbo|]=
\frac1{4\,\pi\,p_0}\,\d(p_0\,{\mp}\,|p_\sbo|)\,\dO^4p~,
\\[6pt]
&e^\pm_0\equiv
\pv\bigl(\frac{\dO^4p}{4\,\pi^2\,|p_\sbo|\,(p_0\mp|p_\sbo|)}\bigr)~,\qquad
e_0\equiv e^+_0 - e^-_0=\frac1{2\,\pi^2}\,\pv\frac{\dO^4p}{p^2}~,
\\[6pt]
&\opp{\sst0}{\pm}{\pm}\equiv
\lim_{\e\to0^+}\frac{\dO^4p}{|p_\sbo|\,(p_0\pm|p_\sbo|\pm\iO\e)}~.
\end{align*}

But there are further aspects,
since we have $0$-mass shells and related generalized densities
in $\DO^*\!\M$ and in $\DO\M$ as well.
The analogous distributions on $\DO\M$ will be denoted by analogous symbols
without any mass label, namely
\begin{align*}
&\ep^\pm\equiv
\frac{\pm1}{4\,\pi\,|x_\sbo|}\,\d(x^0\,{\mp}\,|x_\sbo|)\,\dO^4x\equiv
\frac{\pm1}{4\,\pi\,r}\,\d(t\,{\mp}\,r)\,\dO^4x=
\frac{1}{4\,\pi\,x^0}\,\d(t\,{\mp}\,r)\,\dO^4x~,
\\[6pt]
&e^\pm\equiv 
\pv\bigl(\frac{\dO^4x}{4\,\pi^2\,|x_\sbo|\,(x^0\mp|x_\sbo|)}\bigr)\equiv 
\pv\bigl(\frac{\dO^4x}{4\,\pi^2\,r\,(t\,{\mp}\,r)}\bigr)~,\qquad
e\equiv e^+- e^-=\frac1{2\,\pi^2}\,\pv\frac{\dO^4x}{g(x,x)}~,
\\[6pt]
&\opp{}{\pm}{\pm}\equiv
\lim_{\e\to0^+}\frac{\dO^4x}{|x_\sbo|\,(x^0\pm|x_\sbo|\pm\iO\e)}~.
\end{align*}

A further novelty is that for \hbox{$m=0$}
we can express the generalized functions
\hbox{$\digamma^\pm_{\!\!0}$} explicitely:
\begin{lemma}\label{lemma:digamma0}
We have
$$\digamma^\pm_{\!\!0}(t,r)=2\,\pi^2\,r\,\Bigl[
\pv\Bigl(\frac1{4\,\pi^2\,r\,(t+r)}-\frac1{4\,\pi^2\,r\,(t-r)}\Bigr)
\pm\iO\,\Bigl(\frac{\d(t+r)}{4\,\pi\,r}
-\frac{\d(t-r)}{4\,\pi\,r}\Bigr)\Bigr]$$
that is
$$\digamma^\pm_{\!\!0}\,\dO t\we\Sigma=
2\,\pi^2\,\bigl[e^- -e^+ \mp\iO\,(\ep^+ +\ep^-)\bigr]\equiv
2\,\pi^2\,\bigl[-e \mp\iO\,(\ep^+ +\ep^-)\bigr]~.$$
\end{lemma}\proof
Recalling (\sref{ss:Fourier transforms}) the expressions of
\hbox{$\FO^\pm\bigl(\he(ax)\cdot\exp(\pm\iO ax)\bigr)$}
and using the Euler identities we get
\begin{align*}
&\digamma^+_{\!\!0}(t,r)=\int\limits_{-\infty}^{+\infty}\!\!\dO\t\,
\he(\t)\,\sin(r\,|\t|)\,\eO^{-\iO\,\t\,t}=
\sqrt{2\,\pi}\,\FO\bigl(\he(\t)\,\sin(r\,\t)\bigr)=
\\[6pt]
&\qquad=\frac1{2}\,\bigl(\pv\frac1{t+r}-\pv\frac1{t-r}\bigr)
+\frac{2\,\pi}{4\,\iO}(\d(t-r)-\d(t+r))~,
\end{align*}
and also obtain $\digamma^-_{\!\!0}(t,r)$ as $\digamma^+_{\!\!0}(-t,r)$\,.
The second formula follows immediately from the definitions of the involved objects,
taking \hbox{$r\,\dO t\we\Sigma=\dO^4x$} into account.\qed

In the most important formulas, the generalized densities
$e^\pm$ and $e^\pm_0$ always appear
in the combinations (\sref{ss:Mass-shell related principal values})
$$e\equiv e^+\,{-}\,e^-=
\frac1{2\,\pi^2}\,\pv\bigl(\frac{\dO^4x}{g(x,x)}\bigr)~,\quad
e_0\equiv e^+_0\,{-}\,e^-_0=
\frac1{2\,\pi^2}\,\pv\bigl(\frac{\dO^4\pp}{g^\#(p,p)}\bigr)~.$$
\begin{proposition}\label{proposition:Fopepe0}
We have
\begin{align*}
&\hat\ep^\pm_0=\mp\oh\,e-\ih\,(\ep^+ +\ep^-)~,
&& \check\ep^\pm_0=\mp\oh\,e+\ih\,(\ep^+ +\ep^-)~,
\\[6pt]
&\hat\ep^+_0 +\hat\ep^-_0=-\iO\,(\ep^+ +\ep^-)~,
&& \check\ep^+_0 +\check\ep^-_0=+\iO\,(\ep^+ +\ep^-)~,
\\[6pt]
&\hat\ep^+_0 -\hat\ep^-_0=-e~,
&&\check\ep^+_0 -\check\ep^-_0=-e~,
\\[6pt]
&\hat e_0=\ep^- - \ep^+~, && \check e_0=\ep^- - \ep^+~.
\end{align*}
\end{proposition}\proof
From proposition~\ref{proposition:Fepsilon}
and lemma~\ref{lemma:digamma0} we have
\begin{align*}
&\hat\ep^\pm_0=\frac{\pm1}{(2\pi)^2}\,\digamma^\pm_{\!\!0}\,\dO t\we\Sigma=
\mp\oh\,e-\ih\,(\ep^+ +\ep^-)~,
\\[6pt]
&\check\ep^\pm_0=\frac{\pm1}{(2\pi)^2}\,\digamma^\mp_{\!\!0}\,\dO t\we\Sigma=
\mp\oh\,e+\ih\,(\ep^+ +\ep^-)~.
\end{align*}
The other formulas easily follow from these.
In particular, the last line is obtained by applying $\FO^\mp$
to \hbox{$\FO^\pm\!(\ep^+\,{-}\,\ep^-)=-e_0$}\,,
which is the specular form of the demi-last line.\qed

\remark~We also have the specular identities
\begin{align*}
&\hat\ep^\pm=\mp\oh\,e_0-\ih\,(\ep^+_0 +\ep^-_0)~,
&& \check\ep^\pm=\mp\oh\,e_0+\ih\,(\ep^+_0 +\ep^-_0)~,
\\[6pt]
&\hat\ep^+ +\hat\ep^-=-\iO\,(\ep^+_0 +\ep^-_0)~,
&& \check\ep^+ +\check\ep^-=+\iO\,(\ep^+_0 +\ep^-_0)~,
\\[6pt]
&\hat\ep^+ -\hat\ep^-=-e_0~,
&&\check\ep^+ -\check\ep^-=-e_0~,
\\[6pt]
&\hat e=\ep^-_0 - \ep^+_0~, && \check e=\ep^-_0 - \ep^+_0~,
\end{align*}
obtained by exchanging analogous objects in $\DO\M$ and $\DO^*\!\M$.

\subsubsection{Propagators in the massless case}
\label{sss:Propagators in the massless case}

For \hbox{$m=0$} we deal with solutions and elementary solutions
of the wave equation.
We simplify our notation as \hbox{$\Dcal^\pm\equiv\Dcal^\pm_{\!\sst0}$},
and obtain
$$\Dcal^\pm\eta=-\check\ep^\mp_0=+\hat\ep^\pm_0=
\mp\oh\,e-\ih\,(\ep^+ +\ep^-)~.$$
Moreover we write \hbox{$\Dcal\equiv\Dcal^+ + \Dcal^-$},
\hbox{$\Dcal^\circ\equiv\Dcal^+ - \Dcal^-$},
and obtain the identities
\begin{align*}
&\Dcal\,\eta=
\hat\ep^+_0 +\hat\ep^-_0=-(\check\ep^+_0 +\check\ep^-_0)=
-\iO\,(\ep^+ +\ep^-)=-\tfrac1{2\pi}\,\FO^-\!\om[g^\#(p,p)]~,
\\[6pt]
&\Dcal^\circ\eta=\hat\ep^+_0 -\hat\ep^-_0=\check\ep^+_0 -\check\ep^-_0=-e=
-\tfrac{1}{2\,\pi^2}\,\pv\frac{\eta}{g(x,x)}~.
\end{align*}
Then $\Dcal^\pm\eta$\,, $\Dcal\,\eta$ and $\Dcal^\circ\eta$
are solutions of the homogeneus wave equation.
We also consider the elementary solutions\footnote{
The fact that \hbox{$\iO\,\Dcal^\ret\eta=\ep^+$} and
\hbox{$\iO\,\Dcal^\adv\eta=-\ep^-$}
are elementary solutions of the wave equation,
namely \hbox{$\square\ep^{\pm}=\pm\d$}\,,
can be calso checked by a direct calculation.
See for example Choquet-Bruhat and DeWitt-Morette~\cite{CdW},
\S\,VI.C.5, page~511.} 
\begin{align*}
&\iO\,\Dcal^\bullet\eta\equiv -\oh\,\check e_0\quad\Rightarrow\quad
\Dcal^\bullet\eta=\ih\,\check e_0=-\ih\,(\ep^+ - \ep^-)~,
\\[6pt]
&\iO\,\Dcal^\ret\eta\equiv(\iO\,\Dcal^\bullet+\ih\,\Dcal)\,\eta \quad\Rightarrow\quad
\Dcal^\ret\eta=-\iO\,\ep^+~,
\\[6pt]
&\iO\,\Dcal^\adv\eta\equiv(\iO\,\Dcal^\bullet-\ih\,\Dcal)\,\eta \quad\Rightarrow\quad
\Dcal^\adv\eta=\iO\,\ep^- ~,
\\[6pt]
&\iO\,\Dcal^\Fey\eta\equiv
(\iO\,\Dcal^\bullet+\ih\,\Dcal^\circ)\,\eta \quad\Rightarrow\quad
\Dcal^\Fey\eta=-\oh\,e-\ih\,(\ep^+ - \ep^-)~.
\end{align*}
and get the identities
\begin{align*}
&\oh\,(\Dcal^\ret+\Dcal^\adv)=\Dcal^\bullet~,&&
\Dcal^\ret-\Dcal^\adv=\Dcal~,
\\[6pt]
&\Dcal^\ret=\he(t)\,\Dcal~,&&
\Dcal^\adv=-\he(-t)\,\Dcal~,&&
\Dcal^\Fey=\he(t)\,\Dcal^+-\he(-t)\,\Dcal^-~,
\\[6pt]
&\Dcal^+(-x)=-\Dcal^-(x)~,&&
\Dcal(-x)=-\Dcal(x)~,&&
\Dcal^\circ(-x)=\Dcal^\circ(x)~,
\\[6pt]
&\Dcal^\ret(-x)=\Dcal^\adv(x)~,&&
\Dcal^\bullet(-x)=\Dcal^\bullet(x)~,&&
\Dcal^\Fey(-x)=\Dcal^\Fey(x)~.
\end{align*}

\bigbreak
Finally we observe that, since (\sref{ss:Mass-shell Leray densities})
$$\frac1{2\,\pi}\,\he(\pm x^0)\,\d(g(x,x))\,\eta=
\frac{\pm1}{4\,\pi\,|x_\sbo|}\,\d(x_0\mp|x_\sbo|)=\pm\ep^\pm~,$$
we can also write
\begin{align*}
&\Dcal^\ret(x)=\frac{-\iO}{4\,\pi\,|x_\sbo|}\,\d(x^0-|x_\sbo|)=
-\iO\,\frac{\he(x^0)}{2\,\pi}\,\d(g(x,x))~,
\\[6pt]
&\Dcal^\adv(x)=\frac{-\iO}{4\,\pi\,|x_\sbo|}\,\d(x^0+|x_\sbo|)=
-\iO\,\frac{\he(-x^0)}{2\,\pi}\,\d(g(x,x))~,
\\[6pt]
&\Dcal(x)=\frac{-\iO}{2\,\pi}\,\bigl(\he(x^0)-\he(-x^0)\bigr)\,\d(g(x,x))=
\frac{-\iO}{2\,\pi}\,\sgn(x^0)\,\d(g(x,x))~.
\end{align*}

\remark~The considered solutions and elementary solutions of the wave equation
can be written as combinations of the special generalized densities
$e$ and $\ep^\pm$ on $\DO\M$.
This semplification does not occur in the massive case \hbox{$m\neq0$}.

\subsection{Propagators and vector-valued fields}
\label{ss:Propagators and vector-valued fields}

Elementary solutions are meant to describe the field produced by
a point-like source,
which in the scalar cases on $\DO\M$ examined so far
is situated in \hbox{$\{0\}\in\DO\M$}\,.
In order to transfer our description to the affine Minkowski space,
we consider the identification \hbox{$\M\cong\DO\M$}
determined by any choice of \hbox{$x\in\M$},
so if $\xi$ is an elementary solution we get a kernel
of the form \hbox{$\Xi(x,x')=\xi(x'\,{-}\,x)$}\,.

This procedure can be extended to the case of $\V$-valued fields,
where $\V$ is a vector space,
by starting from $\End\V$-valued distributions on $\DO\M$.
Then the kernel determined by such a distribution is $\End\V$-valued too.
A point-like source is represented by a couple \hbox{$(x,v)\in\M\times\V$},
and $\Xi(x,x')\pint v$ is the corresponding field produced by it.

\subsubsection{Elementary solutions of the Weyl equation}
\label{sss:Elementary solutions of the Weyl equation}

Let $\W$ be the vector space of Dirac spinors and
$$\g^\#:\DO^*\!\M\to\LL^{-1}\tn\End\W:p\mapsto p_\l\g^\l$$
the ``contravariant'' Dirac map (which is a scaled Clifford map). 
A right elementary kernel for $\iO\,\nasl$,
where \hbox{$\nasl\equiv\g^\l\,\na_\l$} is the Dirac operator,
is a generalized map \hbox{$E:\M\times\M\gto\End\W$} fulfilling
(\sref{ss:Elementary solutions of field equations})
$$\iO\,\nasl\!{{\!}_\one}E(x,x')=\Id{\Ww}\tn\d(x-x')~.$$
We look for \hbox{$\Psi\in\LL\tn\DC(\DO\M,\End\W)$} fulfilling
\hbox{$\iO\,\nasl\Psi(x)=\Id{\Ww}\tn\d(x)$}\,,
yielding a right elementary kernel expressed as \hbox{$E(x,x')=\Psi(x'-x)$}\,.
\begin{proposition}
Let $\xi\in\LL^2\tn\SC(\DO\M)$ be an elementary solution
of the wave equation.
Then we get an elementary solution
\hbox{$\Psi:\DO\M\gto\LL\tn\SC(\DO\M,\End\W)$}
of the Weyl equation as
$$\hat\Psi=\g^\#\tn\hat\xi~,$$
namely
$$\Psi=-\iO\,\g^\#\!.\,\xi\equiv -\iO\,\g^\l\tn\de_\l\xi~.$$
Moreover we have
$$\check\Psi=-\g^\#\tn\check\xi~.$$
\end{proposition}
\proof Writing \hbox{$\hat\Psi(p)=(p_\m\,\g^\m)\tn\hat\xi(p)$} we have
$$\FO^+(\iO\,\nasl\Psi)=
-\bigl((p_\l\g^\l)(p_\m\g^\m)\bigr)\tn\hat\xi(p)=
\Id{\Ww}\tn\bigl(-g^{\l\m}\,p_\l\,p_\m\,\hat\xi(p)\bigr)=
\frac1{(2\pi)^2}\,\Id{\Ww}\tn1~.$$
Hence
\hbox{$\iO\,\nasl\Psi\tn\ost\eta=
(2\pi)^{-2}\,\FO^-\bigl(\Id{\Ww}\tn\ost\eta\bigr)
=\Id{\Ww}\tn\d[0]$}\,.
The last formula follows from the expression of the inverse Fourier transform
of a derivative.\qed

Moreover, if \hbox{$\xi$}
is a solution of the homogeneous wave equation then $\g^\#\!.\,\xi$
is obviously a solution of the homogeneous Weyl equation.
If a generalized density which is
a solution or elementary solution of the wave equation
is expressed in terms of the propagator $\Dcal^\square$ as
\hbox{$\xi\equiv\iO\,\Dcal^\square\eta$}\,,
then
$$\Dsl\Dcal{}^\square=-\iO\,\g^\#\!.\Dcal^\square~,\qquad
\FO^+\Dsl\Dcal{}^\square=\g^\#\tn\FO^+\Dcal{}^\square~,\qquad
\FO^-\Dsl\Dcal{}^\square=-\g^\#\tn\FO^-\Dcal{}^\square~.$$

In particular we obtain the solutions
\begin{align*}
&\Dsl\Dcal{}^\pm\eta=\g^\#\!.\bigl(\pm\ih\,e-\oh\,(\ep^+ +\ep^-)\bigr) \quad\Rightarrow\quad
\FO^+(\Dsl\Dcal{}^\pm\eta)=-\g^\#\tn\ep_0^\mp~,\quad
\FO^-(\Dsl\Dcal{}^\pm\eta)=-\g^\#\tn\ep_0^\pm~,
\\[6pt]
&\Dsl\Dcal\,\eta=-\g^\#\!.(\ep^+ + \ep^-) \quad\Rightarrow\quad
\FO^\pm(\Dsl\Dcal\,\eta)=-\g^\#\tn(\ep_0^+ + \ep_0^-)~,
\\[6pt]
&\Dsl\Dcal{}^\circ\eta=\iO\,\g^\#\!.e \quad\Rightarrow\quad
\FO^\pm(\Dsl\Dcal{}^\circ\eta)=\pm\g^\#\tn(\ep_0^+ - \ep_0^-)~,
\end{align*}
and the elementary solutions $\iO\,\Dsl\Dcal{}^\bullet\eta$\,,
$\iO\,\Dsl\Dcal{}^\ret\eta$\,, $\iO\,\Dsl\Dcal{}^\adv\eta$
and $\iO\,\Dsl\Dcal{}^\Fey\eta$\,, where
\begin{align*}
&\Dsl\Dcal{}^\bullet\eta=-\oh\,\g^\#\!.(\ep^+ - \ep^-) \quad\Rightarrow\quad
\FO^+(\Dsl\Dcal{}^\bullet\eta)=-\FO^-(\Dsl\Dcal{}^\bullet\eta)=\ih\,\g^\#\tn e_0~,
\\[6pt]
&\Dsl\Dcal{}^\ret\eta=-\g^\#\!.\ep^+ \quad\Rightarrow\quad
\FO^\pm(\Dsl\Dcal{}^\ret\eta)=\g^\#\tn\bigl(\pm\ih\,e-\oh\,(\ep^+ +\ep^-)\bigr)~,
\\[6pt]
&\Dsl\Dcal{}^\adv\eta=\g^\#\!.\ep^- \quad\Rightarrow\quad
\FO^\pm(\Dsl\Dcal{}^\ret\eta)=\g^\#\tn\bigl(\pm\ih\,e+\oh\,(\ep^+ +\ep^-)\bigr)~,
\\[6pt]
&\Dsl\Dcal{}^\Fey\eta=\g^\#\!.\bigl(\ih\,e-\oh\,(\ep^+ -\ep^-)\bigr) \quad\Rightarrow\quad
\FO^\pm(\Dsl\Dcal{}^\Fey\eta)=\pm\g^\#\tn\bigl(\ih\,e_0+\oh\,(\ep_0^+ -\ep_0^-)\bigr)~.
\end{align*}

Finally we observe that these objects fulfill the same algebraic relations
as the analogous solutions of the wave equation.

\subsubsection{Elementary solutions of the Dirac equation}
\label{sss:Elementary solutions of the Dirac equation}

We look for \hbox{$\Psi\in\LL\tn\DC(\DO\M,\End\W)$} fulfilling
$$\iO\,\nasl\Psi(x)-m\,\Psi(x)=\Id{\Ww}\tn\d(x)~,\quad m\in\LL^{-1}~,$$
yielding a right elementary kernel expressed as \hbox{$E(x,x')=\Psi(x'-x)$}\,.
\begin{proposition}
Let $\xi\in\LL^2\tn\SC(\DO\M)$ be an elementary solution
of the Klein-Gordon equation.
Then we get an elementary solution \hbox{$\Psi:\DO\M\gto\LL\tn\End\W$}
of the Weyl equation by setting
$$\hat\Psi=(\g^\#-m\,\Id{\Ww})\tn\hat\xi~,$$
namely
$$\Psi=-\iO\,\g^\#\!.\,\xi-m\,\xi \equiv (-\iO\,\g^\l\tn\de_\l -m)\xi~.$$
Moreover we have
$$\check\Psi=-(\g^\#+m\,\Id{\Ww})\tn\check\xi~.$$
\end{proposition}
\proof Writing \hbox{$\hat\Psi(p)=(p_\m\,\g^\m-m)\tn\hat\xi(p)$} we have
\begin{align*}
\FO^+(\iO\,\nasl\Psi)&=
-\bigl((p_\l\g^\l+m)(p_\m\g^\m-m)\bigr)\tn\hat\xi(p)=
\\[6pt]
&=\Id{\Ww}\tn\bigl((-g^{\l\m}\,p_\l\,p_\m+m^2)\,\hat\xi(p)\bigr)=
\frac1{(2\pi)^2}\,\Id{\Ww}\tn1
\\[8pt] \Rightarrow\quad&
(\iO\,\nasl-m\,\Id{\Ww})\Psi\tn\ost\eta
=\FO^-\bigl(\tfrac1{(2\pi)^2}\,\Id{\Ww}\tn\ost\eta\bigr)=\Id{\Ww}\tn\d[0]~.
\end{align*}
\qed

If \hbox{$\xi$} is a solution of the homogeneous Klein-Gordon equation
then \hbox{$\iO\,\g^\#\!.\,\xi+m\,\xi$}
is a solution of the homogeneous Dirac equation.
If a generalized density which is
a solution or elementary solution of the Klein-Gordon equation
is expressed in terms of the propagator $\Dcal^\square_{\!m}$ as
\hbox{$\xi\equiv\iO\,\Dcal^\square_{\!m}\,\eta$}\,,
then
$$\Dsl\Dcal{}^\square_{\!m}=-(\iO\,\g^\#\!.+m)\Dcal^\square_{\!m}~,\quad
\FO^+\Dsl\Dcal{}^\square_{\!m}=(\g^\#-m)\tn\FO^+\Dcal{}^\square_{\!m}~,\quad
\FO^-\Dsl\Dcal{}^\square_{\!m}=-(\g^\#+m)\tn\FO^-\Dcal{}^\square_{\!m}~.$$

In particular we obtain the solutions
\begin{align*}
&\Dsl\Dcal{}^\pm_{\!m}\eta=(\iO\,\g^\#\!.+m)\check\ep^\mp_m
\quad\Rightarrow\quad
\FO^+(\Dsl\Dcal{}^\pm_{\!m}\eta)=(-\g^\#+m)\tn\ep^\mp_m~,
\\[6pt]
&\Dsl\Dcal_{\!m}\,\eta=
(\iO\,\g^\#\!.+m)(\check\ep^+_m+\check\ep^-_m) \quad\Rightarrow\quad
\FO^+(\Dsl\Dcal_{\!m}\,\eta)=(-\g^\#+m)\tn(\ep_m^+ + \ep_m^-)~,
\\[6pt]
&\Dsl\Dcal{}^\circ_{\!m}\eta=(\iO\,\g^\#\!.+m)(\check\ep^-_m-\check\ep^+_m)
\quad\Rightarrow\quad
\FO^+(\Dsl\Dcal_{\!m}\,\eta)=(-\g^\#+m)\tn(\ep_m^- - \ep_m^+)~,
\end{align*}
and the elementary solutions $\iO\,\Dsl\Dcal{}^\bullet_{\!m}\eta$\,,
$\iO\,\Dsl\Dcal{}^\ret_{\!m}\eta$\,, $\iO\,\Dsl\Dcal{}^\adv_{\!m}\eta$
and $\iO\,\Dsl\Dcal{}^\Fey_{\!m}\eta$\,, where
\begin{align*}
&\Dsl\Dcal{}^\bullet_{\!m}\eta=\oh\,(\g^\#\!.-\iO\,m)\,\check e_m \quad\Rightarrow\quad
\FO^+(\Dsl\Dcal{}^\bullet_{\!m}\eta)=-\ih\,(\g^\#\,{-}\,m)\tn e_m~,
\displaybreak[2]\\[8pt]
&\Dsl\Dcal{}^\ret_{\!m}\eta=-(\iO\,\g^\#\!.+m)
\bigl(\ih\,\check e_m-\oh\,(\check\ep^+_m+\check\ep^-_m)\bigr) \quad\Rightarrow
\\[6pt] & \hspace{3cm}\Rightarrow\quad
\FO^+(\Dsl\Dcal{}^\ret_{\!m}\eta)=
(\g^\#\,{-}\,m)\tn\bigl(\ih\,e-\oh\,(\ep^+ +\ep^-)\bigr)~,
\displaybreak[2]\\[8pt]
&\Dsl\Dcal{}^\adv_{\!m}\eta=-(\iO\,\g^\#\!.+m)
\bigl(\ih\,\check e_m+\oh\,(\check\ep^+_m+\check\ep^-_m)\bigr) \quad\Rightarrow
\\[6pt] & \hspace{3cm}\Rightarrow\quad
\FO^+(\Dsl\Dcal{}^\adv_{\!m}\eta)=
(\g^\#\,{-}\,m)\tn\bigl(\ih\,e+\oh\,(\ep^+ +\ep^-)\bigr)~,
\displaybreak[2]\\[8pt]
&\Dsl\Dcal{}^\Fey_{\!m}\eta=-(\iO\,\g^\#\!.+m)
\bigl(\ih\,\check e_m+\oh\,(\check\ep^+_m-\check\ep^-_m)\bigr) \quad\Rightarrow
\\[6pt] & \hspace{3cm}\Rightarrow\quad
\FO^+(\Dsl\Dcal{}^\Fey_{\!m}\,\eta)=
(\g^\#\,{-}\,m)\tn\bigl(\ih\,e_0+\oh\,(\ep_0^+ -\ep_0^-)\bigr)~.
\end{align*}

These objects fulfill the same algebraic relations
as the analogous solutions of the Klein-Gordon equation.

\section{Graded commutators of quantum fields}
\label{s:Graded commutators of quantum fields}

If the fields of a classical theory are sections
of a vector bundle \hbox{$\E\onto\M$},
then one can consider a quantized theory in which the fields are sections
of a ``quantum bundle'' \hbox{$\EC\equiv\OC\tn\E\onto\M$},
where $\OC$ is a certain infinite-dimensional $\ZZ_2$-graded algebra
(in the case of a gauge field, that is a connection,
one has to ``fix a gauge'' in order to proceed).
One then finds that the graded commutators of quantum fields
evaluated at different spacetime points are strictly related
to the generalized functions examined
in~\sref{s:Special generalized densities on Minkowski spacetime}.

In this section we sketch the basic involved constructions in flat spacetime,
and show how propagators are generated in a generic setting.
Applications to gauge theories and discussions about extension
to curved spacetime can be find in previous papers~\cite{C12a,C14b,C15a}.

\subsection{Quantum states and operator algebra}
\label{ss:Quantum states and operator algebra}

If \hbox{$m\in\LL^{-1}$} then the future mass-shell
\hbox{$\Pm\equiv\ost\K{}_{\!m}^+\subset\DO^*\!\M$}
can be seen as the space of $4$-momenta of particles of mass $m$\,.
The classical ``internal'' particle structure is described
by a vector bundle \hbox{$\Z\onto\Pm$}\,, either real or complex,
which may be not trivial in general.
We'll be involved with the spaces
$$\ZCo^1\equiv\DCho(\Pm,\Z)~,\qquad\ZCo^{*1}\equiv\DCho(\Pm,\Z^*)~,\qquad
\ZC^1\equiv\DCh(\Pm,\Z)~,\qquad
\ZC^{{*}1}\equiv\DCh(\Pm,\Z^*)$$
of $\Z$-valued and $\Z^*$-valued test semi-densities
and generalized semi-densities
(\sref{ss:Generalized currents and semi-densities}).
Moreover for any \hbox{$n\in\NN$} we set
$$\ZCo^n\equiv\lozenge^n\ZCo^1~,~~\ZCo^{*n}\equiv\lozenge^n\ZCo^{*1}~,~~
\ZC^n\equiv\lozenge^n\ZC^1~,~~\ZC^{*n}\equiv\lozenge^n\ZC^{*1}~,$$
where $\lozenge$ denotes either symmetrized or antisymmetrised tensor product
(respectively for bosons and fermions) and
$$\ZCo\equiv\mathop{\textstyle{\bigoplus}}_{n=0}^\infty\ZCo^{n}~,\qquad
\ZCo^*\equiv\mathop{\textstyle{\bigoplus}}_{n=0}^\infty\ZCo^{*n}~,\qquad
\ZC\equiv\mathop{\textstyle{\bigoplus}}_{n=0}^\infty\ZC^{n}~,\qquad
\ZC^*\equiv\mathop{\textstyle{\bigoplus}}_{n=0}^\infty\ZC^{*n}~.$$

We say that $\ZC$ and $\ZC^*$ are the \emph{multi-particle state spaces}
for a given particle type and the corresponding antiparticle, respectively.
Repeating the above constructions for each one of many particle types,
with masses \hbox{$m',m'',\dots$} and internal bundles $\Z',\Z'',\dots$\,,
we can assemble all of them into total state spaces
$$\VC:=\ZC'\tn\ZC''\tn\mdots
\equiv\mathop{\textstyle{\bigoplus}}_{n=0}^\infty\VC^{n}~,\qquad
\VC^*:=\ZC'^*\tn\ZC''^*\tn\mdots
\equiv\mathop{\textstyle{\bigoplus}}_{n=0}^\infty\VC^{*n}~,$$
where $\VC^{n}$ and $\VC^{*n}$, constituted of all elements of tensor rank $n$\,,
are respectively the spaces of all states of $n$ particles and anti-particles
of any type.
Similarly we construct the dense subspaces \hbox{$\VCo\subset\VC$}
and \hbox{$\VCo^*\subset\VC$} generated by test semi-densities.
Finally we set
$$\QC:=\VC\tn\VC^*~,\qquad \QCo:=\VCo\tn\VCo^*~,$$
which are the \emph{quantum configuration space}
and its subspace generated by test semi-densities.

A monomial element \hbox{$\phi\in\QC$}
may have fermionic and bosonic tensor factors.
We define its \emph{grade} $\grade{\phi}$
to be the number of its fermionic factors
(that is its skew-symmetric tensor algebra rank).
Then we have on $\QC$ a natural structure of $\ZZ_2$-graded algebra,
as we can define the ``exterior product'' of any two elements
by doing either symmetrized or skew-symmetrized tensor products
of the appropriate factors,\footnote{
If (say) \hbox{$\phi'\tn\phi'',\psi'\tn\psi''\in\ZC'\tn\ZC''$}
then \hbox{$(\phi'\tn\phi'')\lozenge(\psi'\tn\psi'')=
(\phi'\lozenge\psi')\tn(\phi''\lozenge\psi'')$}.} 
and extending it by linearity.
Moreover an obvious extension of the contraction between
distributions and test objects yields
an ``interior product'' $(\ze\,|\,\psi)$ between elements \hbox{$\ze\in\VC^*$}
and \hbox{$\psi\in\VC$},
well-defined when either factor is a test object
and, actually, well-defined in many more cases
though \emph{not} for \emph{any} two factors.
If \hbox{$\ze\in\VC^{*1}$} and \hbox{$z\in\VC^1$}
then \hbox{$(\ze\,|\,z)\equiv\bbang{\ze,z}$}.
Whenever the involved factors have definite grade and all expressions
are well-defined, we get the rules
\begin{align*}
&\psi\swe\phi=(-1)^{\grade{\phi}\grade{\psi}}\phi\swe\psi~,\quad
(\ze\swe\xi)\,|\,\psi=\xi\,|\,(\ze\,|\,\psi)~,
\\[6pt]
&\ze\,|\,(\phi\swe\psi)=
(\ze\,|\,\phi)\swe\psi+(-1)^{\grade{z}\grade{\phi}}\,\phi\swe(\ze\,|\,\psi)~,
\qquad\phi,\psi\in\VC,~\ze,\xi\in\VC^{{*}1},
\end{align*}
which are essentially the same rules as one has in finite-dimensional
exterior algebra.

For \hbox{$\ze\in\VC^{{*}1}$} and \hbox{$z\in\VC^1$} we have the linear maps
$$\abs{a}[\ze]:\VCo\to\VCo:\phi\mapsto\abs{a}[\ze]\phi\equiv (\ze\,|\,\phi)~,\quad
\tra{a}[z]:\VCo\to\VC:\phi\mapsto\tra{a}[z]\phi\equiv z\swe\phi~,$$
respectively called the \emph{absorption} operator associated with $\ze$
and the \emph{emission} operator associated with $z$.
Similarly, we have operators
\hbox{$\abs{a}[z]:\VCo^{*}\to\VCo^{*}$} and
\hbox{$\tra{a}[\ze]:\VCo^{*}\to\VC^{*}$},
and one easily checks that $\abs{a}[\ze]$ and $\tra{a}[\ze]$
are mutually transposed maps.

Let now the \emph{grades} of $\abs{a}[z]$ and $\tra{a}[\ze]$
be respectively $\grade{z}$ and $\grade{\ze}$\, and define the
\emph{graded commutator} of any two graded operators as
$$\suc{X,Y}:=X\,Y-(-1)^{\sst\grade{X}\grade{Y}}Y\,X~.$$
Then the above algebraic rules yield
$$\suc{\abs{a}[\xi],\abs{a}[\ze]}=\suc{\tra{a}[y],\tra{a}[z]}=0~,\quad
\suc{\abs{a}[\ze],\tra{a}[z]}=\bbang{\ze,z}\,\id~.$$
We now observe that the composition \hbox{$\abs{a}[z]\comp\tra{a}[\ze]$}
is not a well-defined operation for any $z$ and $\ze$,
while \hbox{$\tra{a}[\ze]\comp\abs{a}[z]$} is always well-defined.
More generally any composition of $n$ emission and absorption operators,
arranged in such a way that all absorption operators stand on the right
of any emission operators, is a well-defined linear operator \hbox{$\QCo\to\QC$}.
Let $\OC^n$ the vector space of all linear combinations of such compositions;
we can define a product \hbox{$\OC^n\times\OC^p\to\OC^{n+p}$}
as composition together with \emph{normal reordering},
namely we move absorption operators to the right of emission operators
using the rule
$$\abs{a}[z]\comp\tra{a}[\ze]~\longrightarrow~
(-1)^{\grade{z}\,\grade{\ze}}\,\tra{a}[\ze]\comp\abs{a}[z]~,$$
which amounts to modify the above graded commutator as
\hbox{$\suc{\abs{a}[\ze],\tra{a}[z]}=0$}\,.
In this way, the vector space \hbox{$\OC'\equiv\bigoplus_{n=0}^\infty\OC^n$}
acquires a structure of $\ZZ_2$-graded algebra.

Composition \emph{without} normal reordering is also considered,
but it must be intended in a generalized sense
(\sref{ss:Generalized bases and basic operators}).

\subsection{Generalized bases and basic operators}
\label{ss:Generalized bases and basic operators}

We use a chosen orthogonal splitting of $\DO\M$ and of $\DO^*\!\M$
as described at the beginning
of~\sref{s:Special generalized densities on Minkowski spacetime},
and recall that for any mass \hbox{$m\in\LL^{-1}$}
one gets a diffeomorphism \hbox{$\Pm\cong(\DO^*\!\M)^\sbot$}.
Thus the pull-back of the spacelike volume form \hbox{$\ost\eta_\sbo=\dO^3p_\sbo$}
can be viewed as a (scaled) volume form on $\Pm$\,,
which we denote by the same symbol.
It will be often convenient to label momenta \hbox{$p\in\Pm$}
by their spatial part $p_\sbo$\,.

If \hbox{$q\in\Pm$} then the Dirac density \hbox{$\d[q]\in\DC(\Pm,\Z)$}
can be written as \hbox{$\d(p_\sbo\!{-}q_\sbo)\,\dO^3p_\sbo$}\,,
where $\d(p_\sbo\!{-}q_\sbo)$ is regarded as an $\LL^3$-scaled generalized function
of the variable $p$\,.
We now introduce a generalized semi-density
$$\Xsf_q:=l^{-3/2}\,\d(p_\sbo{-}q_\sbo)\,\sqrt{\dO^3p_\sbo}~,$$
where the constant \hbox{$l\in\LL$} is needed
in order to get an unscaled object.

Let now $\bigl(\bb_\a\bigr)$ be a frame of the vector bundle \hbox{$\Z\onto\Pm$}\,,
and consider the set
$$\bigl\{\Bsf_{p\a}\bigr\}\equiv\bigl\{\Xsf_p\tn\bb_\a(p)\bigr\}~,$$
which we call a \emph{generalized basis} of \hbox{$\ZC^1\equiv\DCh(\Pm,\Z)$}\,.
Indeed, it can be shown\footnote{
See Schwartz~\cite{Sc}, Ch.\;3, p.\;75.} 
that the space of all linear combinations of such elements is dense in $\ZC^1$.
Moreover we indicate the dual frame of \hbox{$\Z^*\onto\Pm$}
as $\bigl(\bb^\a\bigr)$,
and set \hbox{$\Xsf^p\equiv\Xsf_p$}\,.
Then we regard the set
$$\bigl\{\Bsf^{p\a}\bigr\}\equiv\bigl\{\Xsf^p\tn\bb^\a(p)\bigr\}$$
as the ``dual'' generalized basis of \hbox{$\ZC^{*1}\equiv\DCh(\Pm,\Z^*)$}\,.
Traditionally, our $\Bsf_{p\a}$ would be rather indicated as $\ket{p,\a}$.
However our conventions yield a handy ``generalised index'' notation.
Though contraction of any two distributions is not defined in the ordinary sense,
we may write
$$\bang{\Bsf^{p'\!\a'},\Bsf_{p\a}}=\d^{p'}_p\,\d^{\a'}_{\a}~,$$
where $\d^{p'}_p$ is the generalised function
usually indicated as $\d(p'_\sbo\,{-}\,p_\sbo)$\,.
This is consistent with ``index summation'' in a generalised sense:
if \hbox{$z\in\ZCo^1$} and \hbox{$\ze\in\ZCo^{*1}$} are test semi-densities,
then we write
\begin{align*}
&z=z^{p\a}\,\Bsf_{p\a}~,\quad
z^{p\a}\equiv z^\a(p)\equiv\bang{\Bsf^{p\a},z}~,
\\[6pt]
&\ze=\ze_{p\a}\,\Bsf^{p\a}~,\quad
\ze_{p\a}\equiv \ze_\a(p)\equiv\bang{\ze,\Bsf_{p\a}}~,
\\[6pt]
&\bang{\ze,z}\equiv
\ze_{p'\!\a'}\,z^{p\a}\,\bang{\Bsf^{p'\!\a'},\Bsf_{p\a}}\equiv
\int_{\Pm} \ze_\a(p)\,z^\a(p)\,\dO^3p_\sbo~,
\end{align*}
namely we interpret index summation
with respect to the continuous variable $x$ as integration,
provided by the chosen volume form.
This formalism can be extended to the contraction
of two generalised semi-densities whenever it makes sense.

In particular we write \hbox{$\abs{a}^{p\a}\equiv\abs{a}[\Bsf^{p\a}]$}\,,
\hbox{$\tra{a}_{p\a}\equiv\tra{a}[\Bsf_{p\a}]$}\,,
and obtain super-commutation rules
$$\Suc{\abs{a}^{p\a},\abs{a}^{p'\!\a'}}=
\Suc{\tra{a}_{p\a},\tra{a}_{p'\!\a'}}=0~,\quad
\Suc{\abs{a}^{p\a},\tra{a}_{p'\!\a'}}=\d^\a_{\a'}\,\d^p_{p'}~,$$
where the latter is to be understood in a generalised sense:
for \hbox{$\ze\in\VCo^{{*}1}$}, \hbox{$z\in\VCo^1$}\,,
we write \hbox{$\abs{a}[\ze]=\ze_{p\a}\abs{a}^{p\a}$},
\hbox{$\tra{a}[z]=z^{p\a}\tra{a}_{p\a}$}\,, and
$$\Suc{\abs{a}[\ze],\tra{a}[z]}=\ze_{p\a}\,
\Suc{\ze_{p\a}\,\abs{a}^{p\a},z^{p'\!\a'}\,\tra{a}_{p'\!\a'}}=
\ze_{p\a}\,z^{p'\!\a'}\,\Suc{\abs{a}^{p\a},\tra{a}_{p'\!\a'}}=
\bang{\ze,z}~.$$

\subsection{Free quantum fields}
\label{ss:Free quantum fields}

As hinted at the beginning of~\sref{s:Graded commutators of quantum fields},
if \hbox{$\E\onto\M$} is a classical vector bundle
then we can view \hbox{$\EC\equiv\OC\tn\E\onto\M$} as the corresponding
``quantum bundle'', whose sections are the quantum fields.
There are various issues involved:
in particular, the construction of $\OC$ depends on the choice
of an observer (determining the orthogonal splitting of $\DO^*\!\M$),
and the possible extension to curved spacetime is not at all obvious.
These questions were discussed to some extent
in previous papers~\cite{C12a,C14b}.
Eventually one must conclude 
that such difficulties cannot be thoroughly and definitely solved
in the context of the existing theory.
A further complication is that the fields, in some sectors,
may be subjected to contraints of various types.
However we maintain that
the setting offered in~\sref{ss:Quantum states and operator algebra}
and~\sref{ss:Generalized bases and basic operators}
is an adequate starting point for an outline of essential notions.

We now summarize some points from the previous sections
and make a few preliminary observations.
Each ``internal'' classical bundle \hbox{$\Z\onto\Pm$},
together with its dual (anti-particle) bundle \hbox{$\Z^*\onto\Pm$},
determines a ``sector'' of the theory under consideration.
The basic operators $\abs{a}^{p\a}$ and $\tra{a}_{p\a}$
are interpreted as absorption and emission of a particle of momentum
\hbox{$p\in\Pm$} and internal state $\bb_\a(p)$\,---\,%
in general the frame depends on momentum.
Similarly we have basic operators
\hbox{$\abs{a}_{p\a}\equiv\abs{a}[\Bsf_{p\a}]$}
and \hbox{$\tra{a}{}^{p\a}\equiv\tra{a}[\Bsf^{p\a}]$}\,,
interpreted as absorption and emission of an anti-particle of momentum
\hbox{$p\in\Pm$} and internal state $\bb^\a(p)$\,.

Next we observe that the above basic operators can be regarded
as \emph{$\OC$-valued generalized maps} on $\Pm$\,,
namely we can write
$$\abs{a}^\a:\Pm\gto\OC:p\mapsto\abs{a}^\a(p)\equiv\abs{a}^{p\a}$$
and the like.
Furthermore they can be naturally regarded as components of sections
$$\abs{a}^\a\,\bb_\a\,,\,\tra{a}{}^\a\,\bb_\a:\Pm\to\OC\tn\Z~,\quad
\abs{a}_\a\,\bb^\a\,,\,\tra{a}_\a\,\bb^\a:\Pm\to\OC\tn\Z^*~,$$
with
\hbox{$\abs{a}^\a\,\bb_\a:p\mapsto\abs{a}^\a(p)\tn\bb_\a(p)$}
and the like.

Since $4$-momenta can be labeled by their spatial part $p_\sbo$\,,
we can select the fiber $\Z_{\!0}$ at \hbox{$p_\sbo=0$}
and represent a particle's internal states as elements in it
provided that we avail of a family of linear isomorphisms
\hbox{$K(p):\Z_{\!0}\to\Z_{\!p}$}\,, \hbox{$p\in\Pm$}\,.
In the main cases of physical interest we actually have this;
moreover the fibers of \hbox{$\Z\onto\Pm$} are endowed with a scalar product,
either real or Hermitian, and $K(p)$ is unitary.
Then we define a \emph{free quantum field} as
\hbox{$\phi\equiv\phi^\a\,\bb_\a(0):\DO\M\to\OC\tn\Z_{\!0}$} where
$$\phi^\a(x)\equiv
\frac1{(2\pi)^{3/2}}\int\frac{\dO^3 p_\sbo}{\sqrt{2\,p_0}}\,
K^\a_\b(p)\bigl(\eO^{-\iO\,\bang{p,x}}\,\abs{a}^\b(p)
+\eO^{\iO\,\bang{p,x}}\,\tra{a}{}^\b(p)\bigr)~,\quad
p_0=\Eo_m(p)~,$$
which, because of the condition \hbox{$p_0=\Eo_m(p)\equiv(m^2+p_\sbo^2)^{1/2}$},
is a combination of a Fourier transform
and a Fourier anti-transform of distributions with support in $\Pm$\,.
Note that in order to get a field on $\M$ one still has to fix a point \hbox{$o\in\M$}
(an ``origin''), determining an identification \hbox{$\DO\M\cong\M$}.
Together with point interactions,
free fields can be regarded as ``building blocks'' of field dynamics.

\remark~Strictly speaking, $\phi^\a$ is valued in the completion of $\OC$
constituted by equivalence classes of curves \hbox{$Z:\RR\to\OC$}
with the following property:
the limit \hbox{$\lim_{\l\to0}[Z(\l)\chi]\in\QC$}
exists in the sense of distributions
for all \hbox{$\chi\in\QCo$}\,.\smallbreak

Now we can just replace $\Z$ with $\Z^*$
and repeat the same construction in order to obtain an anti-particle free field 
\hbox{$\td\phi\equiv\td\phi_\a\,\bb^\a(0):\DO\M\to\OC\tn\Z^*_{\!0}$}\,.
There is one slight complication, however:
in order to eventually obtain the graded commutator identities expected
in quantum field theory we have to set
$$\td\phi_\a(x)=
\frac1{(2\pi)^{3/2}}\int\frac{\dO^3 p}{\sqrt{2\,p_0}}\,
\cev K{}^\b_\a\,(p_\sbo)\Bigl( \pm
\eO^{-\iO\,\bang{p,x}}\,\abs{a}_\b(p_\sbo)
+\eO^{\iO\,\bang{p,x}}\,\tra{a}_\b(p_\sbo) \Bigr)~,\quad
p_0=\Eo_m(p)~,$$
where the double sign refers to bosonic and fermionic fields, respectively.

\remark~If $\Z$ is complex then a Hermitian scalar product on its fibers
determines an isomorphism \hbox{$\Z^*\cong\Zc$},
where the latter is the \emph{conjugate bundle}.
Accordingly one can equivalently introduce the field $\td\phi$
as a construction in terms of $\Zc$.
On the other hand,
in the physics literature one often deals with an essentially matrix language,
in which several details of the underlying algebraic structure
are not made explicit,
but we maintain that a few hair-splitting observations can clarify
various issues related to notations and conventions.
\smallbreak

\remark~The case of fields related to Dirac fermions is somewhat more complicate
as the internal structure of a particle and its anti-particle are described
by mutually orthogonal sub-bundles of dual bundles~\cite{C07,C12a,C14b}.
\smallbreak

\subsection{Propagators from graded commutators}
\label{ss:Propagators from graded commutators}

The basic super-commutation rules of emission and absorption operators
can be rewritten as
$$\suc{\abs{a}^\a(p_\sbo)\,,\,\tra{a}_\b(q_\sbo)}=
\suc{\abs{a}_\b(p_\sbo)\,,\,\tra{a}{}^\a(q_\sbo)}=
\d^\a_\b\,\d(p_\sbo-q_\sbo)~,$$
while other super-commutators vanish, namely
\begin{align*}
0&=\suc{\abs{a}^\a(p_\sbo)\,,\,\abs{a}^\b(q_\sbo)}=
\suc{\tra{a}{}^\a(p_\sbo)\,,\,\tra{a}{}^\b(q_\sbo)}=
\suc{\abs{a}_\a(p_\sbo)\,,\,\abs{a}_\b(q_\sbo)}=
\suc{\tra{a}_\a(p_\sbo)\,,\,\tra{a}_\b(q_\sbo)}={}
\\[6pt]
&=\suc{\abs{a}^\a(p_\sbo)\,,\,\abs{a}_\b(q_\sbo)}=
\suc{\tra{a}{}^\a(p_\sbo)\,,\,\tra{a}_\b(q_\sbo)}=
\suc{\abs{a}_\a(p_\sbo)\,,\,\tra{a}_\b(q_\sbo)}=
\suc{\abs{a}^\a(p_\sbo)\,,\,\tra{a}{}^\b(q_\sbo)}~.
\end{align*}

Then, for any two \hbox{$x,x'\in\DO\M$}
we have the vanishing super-commutators
\begin{align*}
&\Suc{\phi^\a(x)\,,\,\phi^\b(x')}=\Suc{\phi^\a(x)\,,\,\phi^{\b{*}}(x')}=
\Suc{\phi^{\a{*}}(x)\,,\,\phi^{\b{*}}(x')}=0~,
\\[6pt]
&\Suc{\phi^\a(x)\,,\,\phi^\b_{,\l}(x')}=
\Suc{\phi^\a(x)\,,\,\phi^{\b{*}}_{,\l}(x')}=
\Suc{\phi^{\a{*}}(x)\,,\,\phi^{\b{*}}_{,\l}(x')}=0~.
\end{align*}
Moreover we find the graded commutators
\begin{align*}
&\Suc{\phi^\a(x)\,,\,\td\phi_\b(x')}=\d^\a_\b\,\Dcal_{\!m}(x-x')~,
\\[6pt]
&\Suc{\phi^\a(x)\,,\,\td\phi_{\b,\l}(x')}=-\d^\a_\b\,\Dcal_{\!m,\l}(x-x')~,
\\[6pt]
&\Suc{\phi^\a_{,\l}(x)\,,\,\td\phi_\b(x')}=\d^\a_\b\,\Dcal_{m\!,\l}(x-x')~,
\end{align*}
where \hbox{$\td\phi_{\a,\l}\equiv\de\td\phi_\a/\de x^\l$},
\hbox{$\phi^\a_{,\l}\equiv\de\phi^\a/\de x^\l$}
(we'll show some explicit calculations below).
We note that these expressions depend on the difference $x\,{-}\,x'$.
This is meaningful, as we recall that in the definitions
of $\phi^\a(x)$ and \hbox{$\td\phi(x)$}
one must have \hbox{$x\in\DO\M$},
so that the choice of a reference point is needed in order
to get fields defined on $\M$.
The above graded commutators, however, turn out to be independent of that choice.

\smallbreak
We are specially interested in \emph{equal times} graded commutators
(\hbox{$x^0=x'^0$}). We get
\begin{align*}
&\Suc{\phi^\a(x)\,,\,\td\phi_\b(x')}=0~,
\\[6pt]
&\Suc{\phi^\a(x)\,,\,\td\phi_{\b,0}(x')}=
-\Suc{\phi^\a_{,0}(x)\,,\,\td\phi_\b(x')}=
-\iO\,\d^\a_\b\,\d(x_\sbo\,{-}\,x'_\sbo)~.
\end{align*}
As already observed (\sref{ss:Klein-Gordon propagators})
the fact that $\Dcal_{\!m}$ is a geometrically well-defined distribution,
independent of the chosen observer,
implies that the above graded commutators vanish whenever \hbox{$x\,{-}\,x'$}
lies outside the causal cone in $\DO\M$.
This fact is important because excludes superluminal influences
in field propagation.

In the generic field theory that we are considering, the most natural
free-field Lagrangian density has an expression of the type
$$\ell\free[\phi,\td\phi]=\bigl(g^{\l\m}\,\td\phi_{\a,\l}\,\phi^\a_{,\m}
-m^2\,\td\phi_\a\,\phi^\a\bigr)\,\dO^4x$$
whence we get the ``conjugate momenta''
\hbox{$\p_\a=\td\phi_{\a,0}$} and \hbox{$\p^\a=\phi^\a_{,0}$}\
respectively associated with $\phi^\a$ and $\td\phi_\a$\,.
We then see that the equal-time graded commutation rules
\begin{align*}
&\Suc{\phi^\a(x)\,,\,\p_\b(x')}=\pm\iO\,\d^\a_\b\,\d(x_\sbo\,{-}\,x'_\sbo)~,
\\[6pt]
&\Suc{\phi^\a(x)\,,\,\phi^\b(x')}=\Suc{\p_\a(x)\,,\,\p_\b(x')}=0~,
\end{align*}
are fulfilled.
These are required to hold as an implementation of the principle of correspondence
in a Hamiltonian context.
Their validity for interacting fields
(solutions of the full field equations with interactions among different sectors)
can then be inferred by general arguments based on the form of the dynamics.
Note that, in standard expressions written in terms of field components,
the product of field components valued at the same spacetime point
is defined through normal ordering
(\sref{ss:Quantum states and operator algebra}),
in order to obtain $\OC$-valued quantities.
Instead, normal ordering is \emph{not} assumed in the above rules,
which must be intended in a generalized distributional sense.

\subsubsection*{Derivation of the graded commutator identities}
We show explicit calculations of the graded commutators
between fields evalued at different points. We have
\begin{align*}
&\Suc{\phi^\a(x)\,,\,\td\phi_\b(x')}=
\frac1{(2\pi)^3}\,\int\frac{\dO^3p_\sbo\,\dO^3q_\sbo}{\sqrt{4\,p_0\,q_0}}\,
K^\a_{\a'}(p_\sbo)\,\cev K{}^{\b'}_\b(q_\sbo)\cdot{}
\\[6pt]
&\qquad\qquad\cdot\Suc{\eO^{-\iO\,\bang{p,x}}\,\abs{a}^{\a'}(p_\sbo)
+\eO^{+\iO\,\bang{p,x}}\,\tra{a}{}^{\a'}(p_\sbo)\,,\,
\pm\eO^{-\iO\,\bang{q,x'}}\,\abs{a}_{\b'}\!(q_\sbo)
+\eO^{+\iO\,\bang{q,x'}}\,\tra{a}_{\b'}\!(q_\sbo)}=
\displaybreak[2]\\[8pt]
&\qquad=\frac1{(2\pi)^3}\,\int\frac{\dO^3p_\sbo\,\dO^3q_\sbo}{\sqrt{4\,p_0\,q_0}}\,
K^\a_{\a'}(p_\sbo)\,\cev K{}^{\b'}_\b(q_\sbo)\cdot{}
\\[6pt]&
\qquad\qquad\cdot\Bigl(
\pm\eO^{\iO\,(-px-qx')}\,\suc{\abs{a}^{\a'}(p_\sbo)\,,\,\abs{a}_{\b'}(q_\sbo)}
+\eO^{\iO\,(-px+qx')}\,\suc{\abs{a}^{\a'}(p_\sbo)\,,\,\tra{a}{}_{\b'}(q_\sbo)}+{}
\\[6pt]&\qquad\qquad\qquad
\pm\eO^{\iO\,(px-qx')}\,\suc{\tra{a}{}^{\a'}(p_\sbo)\,,\,\abs{a}_{\b'}(q_\sbo)}
+\eO^{\iO\,(px+qx')}\,\suc{\tra{a}{}^{\a'}(p_\sbo)\,,\,\tra{a}_{\b'}(q_\sbo)}\Bigr)=
\displaybreak[2]\\[8pt]
&\qquad=\frac1{(2\pi)^3}\,\int\frac{\dO^3p_\sbo\,\dO^3q_\sbo}{\sqrt{4\,p_0\,q_0}}\,
K^\a_{\a'}(p_\sbo)\,\cev K{}^{\b'}_\b(q_\sbo)\,\d^{\a'}_{\b'}\,
\d(p_\sbo-q_\sbo)\,\bigl(\eO^{\iO\,(-px+qx')}-\eO^{\iO\,(px-qx')}\bigr)=
\displaybreak[2]\\[8pt]
&\qquad=\frac1{(2\pi)^3}\,\d^\b_\a\,\int\frac{\dO^3p}{2\,p_0}\,
\bigl(\eO^{\iO\,p\,(-x+x')}-\eO^{\iO\,p\,(x-x')}\bigr)=
\d^\a_\b\,\bigl(\Dcal^+_{\!m}(x-x')+\Dcal^-_{\!m}(x-x')\bigr)~.
\end{align*}

Next we write down the partial derivatives
\begin{align*}
\phi^\a_{,\l}(x)&=\frac\iO{(2\pi)^{3/2}}\int\frac{\dO^3 p}{\sqrt{2\,p_0}}\,
p_\l\,K^\a_\b(p_\sbo)\Bigl(
-\eO^{-\iO\,\bang{p,x}}\,\abs{a}^\b(p_\sbo)
+\eO^{\iO\,\bang{p,x}}\,\tra{a}{}^\b(p_\sbo)\Bigr)~,
\displaybreak[2]\\[8pt]
\td\phi_{\a,\l}(x)&=
\frac\iO{(2\pi)^{3/2}}\int\frac{\dO^3 p}{\sqrt{2\,p_0}}\,
p_\l\,\cev K{}^\b_\a\,(p_\sbo)\Bigl( \mp
\eO^{-\iO\,\bang{p,x}}\,\abs{a}_\b(p_\sbo)
+\eO^{+\iO\,\bang{p,x}}\,\tra{a}_\b(p_\sbo) \Bigr)~,
\end{align*}
and obtain
\begin{align*}
&\Suc{\phi^\a(x)\,,\,\td\phi_{\b,\l}(x')}=
\frac\iO{(2\pi)^3}\,\int\frac{\dO^3p\,\dO^3q}{\sqrt{4\,p_0\,q_0}}\,
p_\l\,K^\a_{\a'}(p_\sbo)\,\cev K{}^{\b'}_\b(q_\sbo)\cdot{}
\\[6pt]
&\qquad\qquad\cdot\Suc{\eO^{-\iO\,\bang{p,x}}\,\abs{a}^{\a'}(p_\sbo)
+\eO^{+\iO\,\bang{p,x}}\,\tra{a}{}^{\a'}(p_\sbo)\,,\,
\mp\eO^{-\iO\,\bang{q,x'}}\,\abs{a}_{\b'}\!(q_\sbo)
+\eO^{+\iO\,\bang{q,x'}}\,\tra{a}_{\b'}\!(q_\sbo)}=
\displaybreak[2]\\[8pt]
&\qquad=\frac\iO{(2\pi)^3}\,\int\frac{\dO^3p\,\dO^3q}{\sqrt{4\,p_0\,q_0}}\,
p_\l\,K^\a_{\a'}(p_\sbo)\,\cev K{}^{\b'}_\b(q_\sbo)\cdot{}
\\[6pt]&
\qquad\qquad\cdot\Bigl(
\mp\eO^{\iO\,(-px-qx')}\,\suc{\abs{a}^{\a'}(p_\sbo)\,,\,\abs{a}_{\b'}(q_\sbo)}
+\eO^{\iO\,(-px+qx')}\,\suc{\abs{a}^{\a'}(p_\sbo)\,,\,\tra{a}_{\b'}(q_\sbo)}+{}
\\[6pt]&\qquad\qquad\qquad
\mp\eO^{\iO\,(px-qx')}\,\suc{\tra{a}{}^{\a'}(p_\sbo)\,,\,\abs{a}_{\b'}(q_\sbo)}
+\eO^{\iO\,(px+qx')}\,\suc{\tra{a}{}^{\a'}(p_\sbo)\,,\,\tra{a}_{\b'}(q_\sbo)}\Bigr)=
\displaybreak[2]\\[8pt]
&\qquad=\frac\iO{(2\pi)^3}\,\int\frac{\dO^3p\,\dO^3q}{\sqrt{4\,p_0\,q_0}}\,
p_\l\,K^\a_{\a'}(p_\sbo)\,\cev K{}^{\b'}_\b(q_\sbo)\,\d^{\a'}_{\b'}\,
\d(p_\sbo-q_\sbo)\,\bigl(\eO^{\iO\,(-px+qx')}+\eO^{\iO\,(px-qx')}\bigr)=
\displaybreak[2]\\[8pt]
&\qquad=\frac\iO{(2\pi)^3}\,\d^\a_\b\,\int\frac{\dO^3p}{2\,p_0}\,p_\l\,
\bigl(\eO^{\iO\,p\,(-x+x')}+\eO^{\iO\,p\,(x-x')}\bigr)=
-\d^\a_\b\,\bigl(\Dcal^+_{\!m,\l}(x-x')+\Dcal^-_{\!m,\l}(x-x')\bigr)~.
\end{align*}

Quite similarly one finds
\hbox{$\Suc{\phi^\a_{,\l}(x)\,,\,\td\phi_\b(x')}=
\d^\a_\b\,\bigl(\Dcal^+_{\!m,\l}(x-x')+\Dcal^-_{\!m,\l}(x-x')\bigr)$}\,.

\section{Conclusions}
\label{s:Conclusions}

This paper is part of an effort aiming at a clearer formulation
of the fundamental mathematical concepts underlying quantum field theory,
with a particular attention given to the differential geometric aspects.
While most notions presented here
can be found elsewhere in the mathematical literature,
a focused presentation is not easily found.
The physical literature, on the other hand,
tends to introduce mathematical concepts when they are needed,
often aimed at the practical use of formulas
rather than at laying a complete framework.
Hence we hope that this paper may fill an actual gap in the existing literature.

Admittedly, here we are not involved with the most advanced concepts
used in recent developments.
However we feel that, in general,
an attitude towards a more focused treatment of the underlying notions
may help in view of overall clarity in every related topic.


\end{document}